\documentclass[article,aps,prd,12pt,notitlepage]{revtex4}

\usepackage{bm}
\usepackage{amssymb}
\usepackage{graphicx,xcolor}
\usepackage{epigraph}
\usepackage{csquotes}
\usepackage{amsmath,amssymb,stmaryrd}
\usepackage{amsmath,graphicx}
\begin{document}
\def\la{{\langle}}
\def\u{\hat U}
\def\U{\hat U}
\def\B{\hat B}
\def\C{\hat C}
\def\D{\hat D}
\def\S{\tilde S}
\def\A{\tilde A}
\def\Delt{\tilde \Delta}
\def\QQ{\hat S}
\def\R{\text {Re}}
\def\e{\enquote}
\def\qq{s}
\def\Q{S}
\def\fb{\overline F}
\def\wb{\overline W}
\def\nl{\newline}
\def\h{\hat H}
\def\ff{\overline q}
\def\k{\overline k}
\def\F {Q}
\def\f{q}
\def\lm{\lambda}
\def\lmu{\underline\lambda}
\def\q{\quad}
\def\t{\tau}
\def\l{\ell}
\def\n{\\ \nonumber}
\def\up{\uppercase}
\def\ra{{\rangle}}
\def\Ep{{\mathcal{E}}}
\def\T{{\mathcal{T}}}
\def\M{{\mathcal{M}}}
\def\omga{{\epsilon}}
\def\t{{\tau}}
\def\h{\hat{H}}
\title{Quantum measurements with, and yet without an Observer }
\author {D. Sokolovski$^{a,b}$} 
\email {dgsokol15@gmail.com}
%\author {A. Matzkin$^c$}
%\author {J. Siewert$^{a,b}$}
%\author {L.M. Baskin$^c$}
%\author {J. G. Muga$^{a,d}$}
\affiliation{$^a$ Departmento de Qu\'imica-F\'isica, Universidad del Pa\' is Vasco, UPV/EHU, Leioa, Spain}
\affiliation{$^b$ IKERBASQUE, Basque Foundation for Science, E-48011 Bilbao, Spain}
%\affiliation{$^c$ Laboratoire de Physique Th\'eorique et Mod\'elisation, CNRS Unit\'e 8089, CY Cergy Paris Universit\'e,
%95302 Cergy-Pontoise cedex, France}
\begin{abstract}
\noindent
\textbf{It is argued that Feynman's rules for evaluating probabilities, combined with 
von Neumann's principle of psycho-physical parallelism, help avoid inconsistencies, often associated 
with quantum theory. The former allows one to assign probabilities to entire sequences of hypothetical Observers' experiences,
without mentioning the problem of wave function collapse.
The latter limits the Observer's (e.g., Wigner's friend's) participation in a measurement to the changes produced in material objects,
thus leaving his/her consciousness outside the picture.}
\end{abstract}
 \date\today
%\pacs{PACS number(s): 03.65.Ta, 73.40.Gk}
%\pacs{03.65.-w, 03.65.Yz, 03.75.Nt}
\maketitle
.
%%%%%%%%%%%%%%
\section{ Introduction} 
Recently there was a renewed interest in in whether quantum theory is internally consistent in its present form, or if new assumptions 
need to be added to its already well established principles. The discussion initiated by the authors of \cite{Renn} was quickly joined, and various opinions were expressed \cite{W1}-\cite{W9}.
An analysis often centres on two issues, the \e{collapse} of the quantum state, and the role and place of a conscious Observer.
The two problems are related. The wave function of the observed system is supposed to undergo a sudden change once 
a definite result of the observation becomes known to the Observer. This change, reminiscent of what happens to a probability 
distribution in classical statistics once additional information is received, may have something to do with Observer's consciousness. 
A related question is how an Observer, taking part in the experiment, should consider other intelligent participants, and whether 
his/her reasoning would depend on availability of the information about other Observers' outcomes \cite{Renn}, or merely on being 
aware of the other measurements being made. One extreme view includes consciousness into a quantum mechanical calculation
directly \cite{LB}, or grants it an active role in the reshaping of the collapsed wave function \cite{Wig}.  On the other extreme, one finds 
theories aiming at denying the Observer any special status at all as happens, for example, in the consistent histories approach (CHA) \cite{W3}, \cite{CHA}.
One cannot help wishing for a compromise position. Would it be possible to have a universal quantum theory centred on the Observer's subjective 
perceptions, and yet applying its mathematical apparatus only to material objects, whenever Observer's probability are calculated? 
One might look for an answer in the literature. 
\newline 
The question was discussed by Bohr \cite{Bohr} and later by von Neumann in his monograph \cite{vN}, both invoking the principle
of psycho-physical parallelism. The principle establishes a correspondence between \e{extra-physical process of subjective perception}
and \e{equivalent physical processes}, as described by the Observer's theory. This is a delicate balancing act. According to 
von Neumann \cite{vN}, its success depends not on providing a detailed explanation of the act of human perception, 
but on being able to move the boundary between \e{physical} and  \e{extra-physical} in an arbitrary manner, deeper into the Observer's body, 
or further out towards the observed system. Von Neumann's discussion covered mostly a single measurement, made on a system 
in an already known quantum state.  
\newline
However, current discussions (see, e.g., \cite{Renn}-\cite{W9}) often consider several consecutive observations, which involve more than one 
Observer. An approach to such situations was outlined in Feynman's undergraduate text \cite{FeynL}, rarely mentioned in the present context.
Feynman's {\it general principles} \cite{FeynL} are quite simple. To find the probability of a sequence of observed events, 
one needs to evaluate the amplitude for each route, by multiplying the amplitudes for each part of the route, add up the amplitudes,
if the routes cannot be told apart, and take the absolute square of the sum. Feynman warns against thinking {\it \e{in terms of \e{particle
waves}}}, and his recipe does not need to address the "collapse" problem. Nor is the role of consciousness discussed in a great detail.
Sect. 2.6 of \cite{FeynL} hints at the importance of the \e{traces left} by a phenomenon, and leaves the problem in that form. 
\newline
The purpose of this paper is to establish whether the principles of \cite{vN} and \cite{FeynL} are sufficient 
to make an intelligent Observer a client and a beneficiary of quantum theory, at the same time keping the subject of consciousness
 outside the theory's scope. 
We will also ask whether, with the first task achieved, the theory is able to provide unequivocal answers in the situations 
where its consistency is questioned. 
\newline
In Sect. II we adopt Feynman's recipe \cite{FeynL} to describe a series of consecutive quantum measurements.
In Sect. III we demonstrate the equivalence between this \e{static} view, and a \e{evolutionist} picture, in which 
an initial quantum state is seen as undergoing a unitary evolution, interrupted by Observers' interventions. 
Sect. IV underlines a distinction between Observer's consciousness, and his/her material memory, thus setting 
a framework for our analysis. 
In Section V we consider the case of two Observers, and three possible scenarios for their experiment. 
In Sect.VI we summarise our preliminary conclusions.
Section VII revisits the Wigner's friend problem of Ref.\cite{Wig}.
In Sect. VIII we discuss an interference experiment, similar to that proposed in \cite{Deu}.
In Sect. IX we show how von Neumann boundary can be placed \e{at the level of the observed system} in a general case. 
Sect. X describes a more efficient way to calculate the probabilities, in which all but the last Observers are represented by their unobserved probes.
In Sect. XI we discuss certain similarities and differences between our analysis, and the consistent histories approach of \cite{CHA}.
Sect. XII contains our conclusions. 
%In the previous Section we had to deal with two types of paths. A virtual path connects basis states, e.g., $|d_m\ra \gets |c_j\ra \gets b_i$. A {\it real} path connects the observed outcomes of the measurements  {\it actually made}, e.g., $D_M\gets C_J\gets B_i$. 
%A real path is a union (superposition) of virtual paths, which cannot be told apart by intermediate made measurements. 
%While a virtual path is always endowed with probability amplitude, a real path has possesses also a probability, with which it is 
%"seen to be travelled". This probability is tColute square of the sum of the amplitudes, belonging to the constituent
%virtual paths. 
%\nl
\section{Quantum rules. The \e{entire history} view.} 
Let us assume that $L+1$ Observers decide to make $L+1$ measurements on particular parts of a quantum system,  with which they associates  a $N$-dimensional Hilbert space. 
If $L+1$ quantities $\mathcal Q^\l$, $\l=0,1,2,...,L$, need to be measured at different times $t=t_\l$, $t_{\l}> t_{\l-1}$,
one associates with each $\mathcal Q^\l$ a discrete orthonormal basis $|q^\l_{n_\l}\ra$, $n_\l=1,...N$, and  a  Hermitian operator, $\hat Q^\l$, whose eigenvalues $Q^\l_{m_\l}$, $m_\l=1...M_\l$ may be degenerate, $M_\l\le N$,
\begin{eqnarray} \label{00}
 \hat Q^\l=\sum_{m_\l=1}^{M_\l} Q^\l_{m_\l}\hat \Pi^\l_{m_\l}, \q \hat \Pi^\l_{m_\l}\equiv \sum_{n_\l=1}^N\Delta\left (Q^\l_{m_\l}-\la q^\l_{n_\l} |\hat Q^\l| q^\l_{n_\l}\ra\right )  |q^\l_{n_\l}\ra\la q^\l_{n_\l}|. 
 \end{eqnarray}
We define $\Delta(x-y)\equiv 1$ for $x=y$, and $0$ otherwise, so that a $\hat \Pi^\l_{m_\l}$ projects onto the eigen-subspace of  the eigenvalue $Q^\l_{m_\l}$. 
Observers' outcomes must coincide with the eigenvalues  of the operators $\hat Q^\l$, and one wishes to evaluate the probabilities 
$P(Q^L_{m_L}...\gets Q^{\l}_{m_{\l}}....\gets Q^0_{m_0})$ of  obtaining a series of outcomes $Q^{L}_{m_{L}}...\gets Q^{\l}_{m_{\l}}...\gets Q^{0}_{m_{0}}$.
 The initial measurement (also known as the preparation) must determine the initial state 
$|q^1_{i_1}\ra$ unambiguously, $Q^0_{m_0} \leftrightarrow |q^0_{n_0}\ra$, and we will always assume that $Q^0_{m_0}$ is non-degenerate. 
\newline
The recipe for constructing the probabilities $P(Q^L_{m_L}...\gets Q^{\l}_{m_{\l}}....\gets Q^0_{m_0})$ is as follows \cite{FeynL}.
First, one constructs  all {\it virtual}  (Feynman) paths,  $\{q^L_{n_L}...\gets q^\l_{n_\l}...\gets q^0_{n_0}\}$, connecting the eigenstates $|q^\l_{i_\l}\ra$, and ascribes to them probability amplitudes 
\begin{eqnarray} \label{01}
A(q^L_{n_L}...\gets q^\l_{n_\l}\gets q^0_{n_0}) =\prod_{\l=1}^L\la q^\l_{n_\l}|\u(t_\l,t_{\l-1})|q^{\l-1}_{n_{\l-1}}\ra 
%\la q^L_{n_L}|\u(t_L,t_{L-1})|q^{L-1}_{n_{L-1}}\ra
%\times \n
%\times \n
.%..\la q^3_{n_3}|\u(t_3,t_{2})|q^{2}_{n_{2}}\ra
%\n
%\la q^2_{n_2}|\u(t_2,t_{1})|q^{1}_{n_{1}}\ra,
\end{eqnarray}
where $\u(t',t)$ is the system's evolution operator. 
\newline
Then one sums the amplitudes in Eq.(\ref{01}) over the degeneracies of all but the last eigenvalues, thus 
obtaining the \e{real} paths,    $\{q^L_{n_L}...\gets Q^\l_{m_\l}...\gets q^0_{n_0}\}$,  endowed with  the {\it probability amplitudes} 
\begin{eqnarray} \label{02}
\A(q^L_{n_L}...\gets Q^{\l}_{m_{\l}}...\gets q^0_{n_0}) =
\sum_{n_1,,...n_{L-1}=1}^N 
\left [\prod_{\l=1} ^{L-1}
%\q\n
\Delta\left (Q^\l_{m_\l}-\la q^\l_{n_\l}|\hat Q^\l\l|q^\l_{n_\l}\ra\right )\right ]
A(q^L_{n_L}...\gets q^{\l}_{n_{\l}}...\gets q^0_{i_0}),\q
%\la q^L_{n_L}|\u(t_L,t_{L-1})|q^{L-1}_{n_{L-1}}\ra....\la q^3_{n_3}|\u(t_3,t_{2})|q^{2}_{n_{2}}\ra
%\la q^2_{n_2}|\u(t_2,t_{1})|q^{1}_{n_{1}}\ra
\end{eqnarray}
as well as the {\it probabilities} 
\begin{eqnarray} \label{03}
p(q^L_{n_L}...\gets Q^{\l}_{m_{\l}}...\gets q^0_{n_0} )\equiv |\A(q^L_{n_L}...\gets Q^{\l}_{m_{\l}}...\gets q^0_{n_0})|^2.
\end{eqnarray}
Finally, one sums the probabilities in Eq.(\ref{03}) over the degeneracies of the last $\hat Q^L$, to obtain 
\begin{eqnarray} \label{04}
P(Q^L_{m_L}...\gets Q^{\l}_{m_{\l}}....\gets Q^0_{m_0}) = \sum_{n_L=1}^N
\Delta\left (Q^L_{m_L}-\la q^L_{n_L}|\hat Q^L|q^L_{n_L}\ra\right ) p(q^L_{n_L}...\gets Q^{\l}_{m_{\l}}...\gets q^0_{n_0} )
\end{eqnarray}
In general, the situation is non-Markovian - the probability $p(q^L_{n_L}...\gets Q^{\l}_{m_{\l}}...\gets q^0_{n_0} )$
does not factorise into $\prod_{\l=1}^L p(Q^{\l}_{m_{\l}}-Q^{\l-1}_{m_{\l-1}})$, unless all the eigenvalues are non-degenerate, 
$M_\l=N$. For this reason, the amplitude $\A(q^L_{n_L}...\gets Q^{\l}_{m_{\l}}...\gets q^0_{n_0})$ has to refer to the 
{\it entire} experiment, which starts with the preparation at $t=t_0$, and end with the last observation made at $t=t_L$.

Finally, we need to assume that the probabilities (\ref{04})  refer to the Observers' experiences, 
and not  to the statements like \e{a physical quantity has a certain value} \cite{vN}.
The situation should, therefore, be like this. In an experiment, involving several steps, each participant can perceive
one of his/her possible outcomes, $Q^{\l}_{m_{\l}}$. Equations (\ref{01})-(\ref{04}) give a recipe for calculating the likelihoods
of all possible sequences of the perceived outcomes {\it one at a time}. The recipe consists in calculating matrix elements of unitary operators, 
multiplying the results, and adding the products, as appropriate. There is no mention of a \e{state evolving throughout experiment},
neither a need to account for the future development of such a state, after the experiment is finished at $t=t_L$.
One does not need to care about what the participants may think or know about each other. The calculation could be made by an Alice, 
who does not take part in the experiment, and remains in the comfort of her office. Her results will apply to any $L+1$ Observers who may or may not 
communicate with each other, as well as to the same Observer, who performs all $L+1$ measurements single handedly. 
%At the same time, the extra-physical process of subjective perception we must assign equivalent process in the ordinary 
%physical space \cite{vN}. The required correspondence is established in Eqs.(\ref{02})-(\ref{04}).
% Feynman's principles of \cite{FeynL},
However, the problem can also be formulated in a different manner. 
%%%%%%%%%%%%%%%%%%%%%%%%%%%%%%
\section{Quantum rules.  The \e{evolutionist} view}
Equation (\ref{04}) can be written in a more familiar way. Defining a partial evolution operator as 
\begin{eqnarray} \label{011}
\u(Q^{L-1}_{m_{L-1}}...\gets Q^{\l}_{m_{\l}}....\gets q^0_{m_0})\equiv \u(t_L,t_{L-1})
\prod_{\l=1}^{L-1}\hat \Pi^\l_{m_\l}\u(t_\l,t_{\l-1}),
\end{eqnarray}
with a property that 
\begin{eqnarray} \label{011a}
\sum_{m_{L-1}...m_1=1}^{M_{L-1}...M_1}\u(Q^{L-1}_{m_{L-1}}...\gets Q^{\l}_{m_{\l}}....\gets Q^1_{m_1})=\u(t_L,t_{0}),
\q\q\q\q\q\q
\end{eqnarray}
and a projector onto the initial state, 
%\begin{eqnarray} \label{011}
$\hat \rho_0\equiv |q_{n_0}^{0}\ra \la q_{n_0}^{0}|$,
%\end{eqnarray}
one can construct a family of $M_1\times M_2 ...\times M_{L-1}$ density operators (mixed states)
\begin{eqnarray} \label{021}
%\hat \rho(\rho_0)= \sum_{m_{L-1}...m_1}
%\rho(Q^{L-1}_{m_{L-1}}...\gets Q^{\l}_{m_{\l}}....\gets Q^1_{m_1}, \rho_0),\n
%\u^{-1}(Q^{L-1}_{m_{L-1}}...\gets Q^{\l}_{m_{\l}}....\gets Q^1_{m_1}),
\rho(Q^{L-1}_{m_{L-1}}...\gets Q^{\l}_{m_{\l}}....\gets Q^1_{m_1}, \rho_0) \equiv \q\q\q\q\q\q	\n
\u(Q^{L-1}_{m_{L-1}}...\gets Q^{\l}_{m_{\l}}....\gets Q^1_{m_1})\hat \rho_0
\u^{\dagger}(Q^{L-1}_{m_{L-1}}...\gets Q^{\l}_{m_{\l}}....\gets Q^1_{m_1}),
\end{eqnarray}
where $\u^\dagger$ is the hermitian conjugate of $\u$.
Now the probabilities in Eq.(\ref{04}) can be obtained just as well by taking a trace,
\begin{eqnarray} \label{031}
P(Q^L_{m_L}...\gets Q^{\l}_{m_{\l}}....\gets Q^0_{m_0}) =\text{tr}\left [\hat \Pi^L_{m_L}\rho(Q^{L-1}_{m_{L-1}}...\gets Q^{\l}_{m_{\l}}....\gets Q^1_{m_1}, \rho_0)\right ]
\end{eqnarray}
Equation (\ref{031}) remains valid for a more general initial state, 
\begin{eqnarray} \label{041}
\rho_0=\sum_\nu w_\nu |q^\nu_0\ra \la q^\nu_0|, 
\end{eqnarray}
where the system is prepared in a state $|q^\nu_0\ra$ with a probability $ w_\nu$, and $|q^\nu_0\ra$ are any normalised, 
yet not necessarily orthogonal states.
\newline
This is a dynamic picture. In Eq.(\ref{021}) the initial state state (\ref{031}) can be seen as evolving until the time of the last observation, 
yet the evolution is not unitary,
\begin{eqnarray} \label{051}
\u^{\dagger}(Q^{L-1}_{m_{L-1}}...\gets Q^{\l}_{m_{\l}}....\gets Q^1_{m_1})\u(Q^{L-1}_{m_{L-1}}...\gets Q^{\l}_{m_{\l}}....\gets Q^1_{m_1})=\n
\hat \Pi^{1\dagger} _{m_1}(t_1)...\times \hat \Pi^{(L-1)\dagger}_{m_{L-1}}\hat \Pi^{L-1}_{m_{L-1}}...\times \hat \Pi^1_{m_1}(t_1) \ne \hat 1.\q\q\q\q\q\q
\end{eqnarray}
In total, there are $M_1\times M_2 ...\times M_{L-1}$ such evolutions, and for someone who wishes to associate quantum mechanics 
with uninterrupted unitary evolution in Eq.(\ref{011a}),  Eqs.(\ref{031})-(\ref{051}) may present a conceptual problem.
A wave function seen as a substance in continuous flow (\ref{011a}),  decimated each time a conscious Observer makes an enquiry and 
perceives an outcome, 
presents a rather bizarre picture.
\newline
This  problem does not arise in the \e{static} view, outlined at the end of the previous Section.
In the following, we will accept the rules of Sect.II as the basic axioms of quantum theory, and treat 
Eqs.(\ref{031}) - Eqs.(\ref{051}) as their consequences \cite{DSepl1}, which can be derived and used, e.g., for computational convenience. 
%Furthermore, according to the rules of the previous Section the probability in Eqs.(\ref{04}) and (\ref{051}) refer the the outcomes
%perceived by conscious Observers.
%%%%%%%%%%%%%%%%%%%%%%%%%%%%%%%%%%
\section{Consciousness, memory and material records}
In the context of the previous Section, it  is only natural to wonder how an act of perceiving an outcome could succeed in replacing a unitary evolution 
(\ref{011a}) by a non-unitary one in Eq.(\ref{051}). With Observer's consciousness now drawn into the discussion, there are at least two 
possibilities. One can
\newline i)  include Observer's consciousness into the scope of quantum theory  \cite{LB}. This is known to lead to a contradiction with what 
we seem to know about intelligent beings \cite{Wig}.
\newline ii)  exclude consciousness from the analysis completely, reduce its role to that of an external  client, and treat the content of Sect.II as
a rule book, with no obligation to give any explanations. This appears to be in line with the approach outlined in \cite{FeynL}, 
and articulated in more detail in \cite{DSepl1}.
\newline
A further insight can, however, be gained at the cost of making additional assumptions. Consider first an example in which classical physics
is used to determine the trajectory of a tennis ball, after
a tennis player sends it back to the partner's half of the court. The player's consciousness 
is evidently involved -  he or she sees the ball coming, chooses the moment and the angle, and
finally strikes the ball with the racket. There are many aspects clearly outside the remit of classical mechanics, yet mechanics does not fail.
Fortunately, to predict the ball's trajectory one does not need to  understand the mental 
processes which led to the force being exerted. Of the whole complex occurrence classical mechanics requires only the things firmly
 within the theory's scope - the ball's position and velocity, and  the force acting at the time it is being hit.
\newline 
Using this analogy, one can try to limit a complex act of Observer's perception to its consequences in the inanimate {\it material} world. While the details of the act itself are outside the quantum theory's scope, its consequences can be discussed and successfully used in practice. 
In order to do so, we will draw a distinction between Observers consciousness (fully outside our analysis)  and his/her material memory 
(subject to our discussion) \cite{DSepl2}. The act of  registering an outcome will be seen as accompanied by  a change in the state 
of the Observer's memory's, $\M$, i.e., by production of a material record. 
%With this achieved, the Observer will be assumed capable of access the memory directly,
%(by undisclosed means, once or several times), thus becoming aware of the perceived result. 
\newline
Furthermore, quantum theory will be expected to apply to all material objects regardless of their size and complexity.
It will be able, therefore, to treat the change in the memory's state without 
questioning how this change came about, just as classical mechanics does not need to question the chain of the player's decisions, 
%made by the player
 leading to the force being exerted on the ball.
 The appearance of this additional (memory's) degree of freedom, entangled with the system, will enlarge the Hilbert space used 
 in the calculation of Sect.II, alter the degeneracies of the measured eigenvalues, and ultimately give different answers, depending
 on whether an Observer perceives his outcome or not.  
It will also follow that without producing such a record
an act of observation \e{should not count}, just as a movement which misses the ball, or a movement contemplated yet not carried out, 
would have no effect on the ball's trajectory.
\newline 
We will need, therefore,  to distinguish three different developments.  First, an Observer {\it couples} his/her visible probe $D$ to the studied quantum system $S$ at $t=\tau$ (we consider all interactions to be instantaneous, and the system itself invisible to the naked eye). Then, at a $t_r>\tau$, he/she {\it registers}  the state of the probe, 
which produces a record in his/her memory. Later still, at $t_p \ge t_r$ the Observer {\it perceives} the outcome, i.e., becomes aware of the 
impression left in his/her memory.  This can happen at the time of the registration, or at later time, when the Observer consults his/her 
memory again.  
The last step may seem redundant, but is necessary for our analysis.  The probabilities in Sect.II, we recall,  refer to the moments 
the Observers are expected to {\it perceive} their outcomes.
%%%%%%%%%%%%%%%%%%%%%%%%%%%%%%%%%%%%%%%%%%
\section{An example with only two participants}
As an illustration, consider only two Observers, subsequently called F and W, 
a two-level system, $S$, two measuring devices (probes) $D^{F,W}$, visible to both 
Observers, and two sets of individual memories, $\M_F$ and $\M_W$. 
The experiment consists in preparing the composite system in a state $|\Phi_0\ra \equiv|\mu^W_0\ra|\mu^F_0\ra |d^W_0\ra |d^F_0\ra|s_0\ra$ at $t_0=0$. 
At $t=\tau^F>0$  $F$ switches on a coupling which, after a  brief interaction, entangles $D^F$ with $S$,  using a particular basis $|s^F_{1,2}\ra$ according to (for details see the Appendix)
\begin{eqnarray} \label{013}
 |d^F_0\ra|s_0\ra  \to \la s^F_1|\u^S(\tau^F,t_0) |s_0\ra|d^F_1\ra|s^F_1\ra + \la s^F_2|\u^S(\tau^F,t_0)|s_0\ra |d^F_2\ra|s^F_2\ra, 
\end{eqnarray}
where $\u^S(\tau^F,t_0)$ is the system's evolution operator $\la s^F_j|s^F_i\ra=\delta_{ij}$, and $\la d^F_j|d^F_i\ra=\delta_{ij}$. 
Note that the device $D^F$ can be as complex as on wishes, bearing in mind that only two of its states
$|d_1\ra$ and $|d_2\ra$, are involved in the experiment. (For example, an outcome could be a sheet o paper, 
on which a printer may write  $"yes"$ or a $"no"$, but not T.S. Eliot's Fourth Quartet. A similar situation occurs, for example, 
in cold matter physics, where only two states of a complex $Rb$ atom are involved in the cooling experiment \cite{Rb}.) 
\newline
Then at $t=t_r^F> \tau^F$, F \e{registers his result} (we might say \e{looks at the probe}) which, by means beyond our knowledge, 
changes the state of his memory according to  
\begin{eqnarray} \label{023}
|\mu^F_0\ra|d^F\ra   \to \la d^F_1|d^F\ra |\mu^F_1\ra|d^F_1\ra +\la d^F_2|d^F\ra |\mu^F_2\ra|d^F_2\ra, 
\end{eqnarray}
where $|d^F\ra$ is any state of the F's probe,  and $\la \mu^F_j|\mu^F_i\ra=\delta_{ij}$.
After that, at 
\begin{eqnarray} \label{013a}
t_1\equiv t_p^F\ge t_r^F, 
\end{eqnarray}
F accesses his memory (we do not need to know how), and becomes aware of (perceives)  his outcome, a $"yes^F"$ or a $"no^F"$. 
\newline 
Finally, the second Observer, W, is also able to measure the system, F's probe, or their composite, by coupling his own visible device $D^W$ at $t=\tau^W>t^W_r$, 
registering the result at $t_r^W > \tau^W$, and becoming aware of his outcome (perceiving his outcome) $yes^W$ or a $no^W$ upon consulting his memory immediately after, at
\begin{eqnarray} \label{023a}
t_2 \equiv t^W_p=t_r^W+\epsilon, \q \epsilon \to 0.
\end{eqnarray}
For this W will need a probe with four orthogonal states $|d^W_j\ra$, $j=1,2,3,4$, a four-state orthogonal basis for the composite 
{\it system + F's probe}, e.g.,  
\begin{eqnarray} \label{033}
|\phi^W_1\ra=\alpha |s^W_1\ra|d^F_1\ra+\beta |s^W_2\ra|d^F_2\ra, \q |\phi^W_2\ra=\beta^* |s^W_1\ra|d^F_1\ra-\alpha^* |s^W_2\ra|d^F_2\ra,\n
|\phi^W_3\ra=\alpha |s^W_1\ra|d^F_2\ra+\beta |s^W_2\ra|d^F_1\ra, \q |\phi^W_4\ra=\beta^* |s^W_1\ra|d^F_2\ra-\alpha^* |s^W_2\ra|d^F_1\ra,
\end{eqnarray}
where $\la s_j^W|s^W_i\ra=\delta_{ij}$ and $|\alpha|^2+|\beta|^2=1$,
 and a coupling which entangles W's probe with the composite according to
\begin{eqnarray} \label{043}
|d^W_0\ra |\phi \ra  \to \sum_{j=1}^4 \la \phi^W_j|\phi\ra |d^W_j\ra |\phi^W_j\ra, 
\end{eqnarray}
where $|\phi \ra$ is any state of the composite (see Appendix). Finally, W registers the state of his probe according to 
\begin{eqnarray} \label{053}
|\mu^F_0\ra|d^W\ra   \to \sum_{j=1}^4 \la d^W_j|d^W\ra |\mu ^W_j\ra |d^W_j\ra. 
\end{eqnarray}
To apply the rules of Sect.II we note that the acts of coupling, (\ref{013}) and (\ref{043}), and the acts of registering, 
(\ref{023}) and (\ref{053}), must be described by the evolution operators $\u$ in Eq.(\ref{02}), now acting in the Hilbert space
of a composite {\it \{spin +F's probe+ F's memory+ W's probe + W's memory\}}.   
Different outcomes, perceived by F and W, are in one-to-one correspondence with distinct eigenvalues 
of operators $\hat Q^F$ and  $\hat Q^W$, acting in the Hilbert spaces of F's and W's memories, respectively.
Equation (\ref{04}) will then give the probabilities of the possible outcomes, as perceived by the Observers. 
In the following we will assume that the probes and the memories, unlike the system, have no own dynamics, and consider several possible scenarios.  
\newline 
{\it Scenario (A): F does not register his outcome, and W registers and perceives his.} 
If so, F's probe is coupled to the system as in Eq.(\ref{013}), but his memory remains unchanged, since (\ref{023}) has not been applied. 
W's perceived outcomes can be represented, e.g.,  by an operator 
\begin{eqnarray} \label{Wop}
\hat Q^W=|\mu^W_1\ra \la \mu^W_1|+2|\mu^W_2\ra \la \mu^W_2|+
3[|\mu^W_3\ra \la \mu^W_3|+|\mu^W_4\ra \la \mu^W_4|], 
\end{eqnarray}
with the eigenvalues interpreted as $Q^W_1=1\to yes^W$, 
$Q^W_2=2\to no^W$, and $Q^W_3=3\to \{not\q sure\}^W$. 
By this we mean that the probabilities given by Eq.(\ref{04}) for the eigenvalues $Q^W_i$ are the actual odds 
on W saying that his experiment produced an outcome $yes$, $no$, or neither of the two.
% We cannot and do not need to know 
%what, if anything, happens in F's consciousness. What is important to us is that after accepting the above rule, 
%its is possible to continue to apply quantum mechanics. 
With F's memory not involved, and remaining in its initial state $|\mu^F_0\ra$, the basis states of the joint system
{\it \{system+F's probe+W's probe+W's memory \}}
are conveniently chosen as 
\begin{eqnarray} \label{Wst}
|q^W_{ijk}\ra=|\mu^W_k\ra|d^W_j\ra|\underline{ \phi} ^W_i\ra, \q i,j,k=1,2,3,4,
\end{eqnarray}
where $|\underline \phi^W_i\ra\equiv \u^S(t^W_p,\tau^W)| \phi^W_i\ra$ and $\u^S(t^W_p,\tau^W)$ is the system's evolution operator. We will need the matrix elements, $\la q^W_{ijk}|\U(t^W_p,t_0)|\mu^W_0\ra|d^W_0\ra |d^F_0\ra|s_0\ra$, where $\U(t^W_p,t_0)$ takes into account the developments (\ref{013}), (\ref{043}), and (\ref{053}), but not (\ref{023}).
 The matrix elements are easily evaluated, and we find
%%%%%%%%%%%%%%%%%%%
\begin{figure}[h]
\includegraphics[angle=0,width=16cm, height= 8cm]{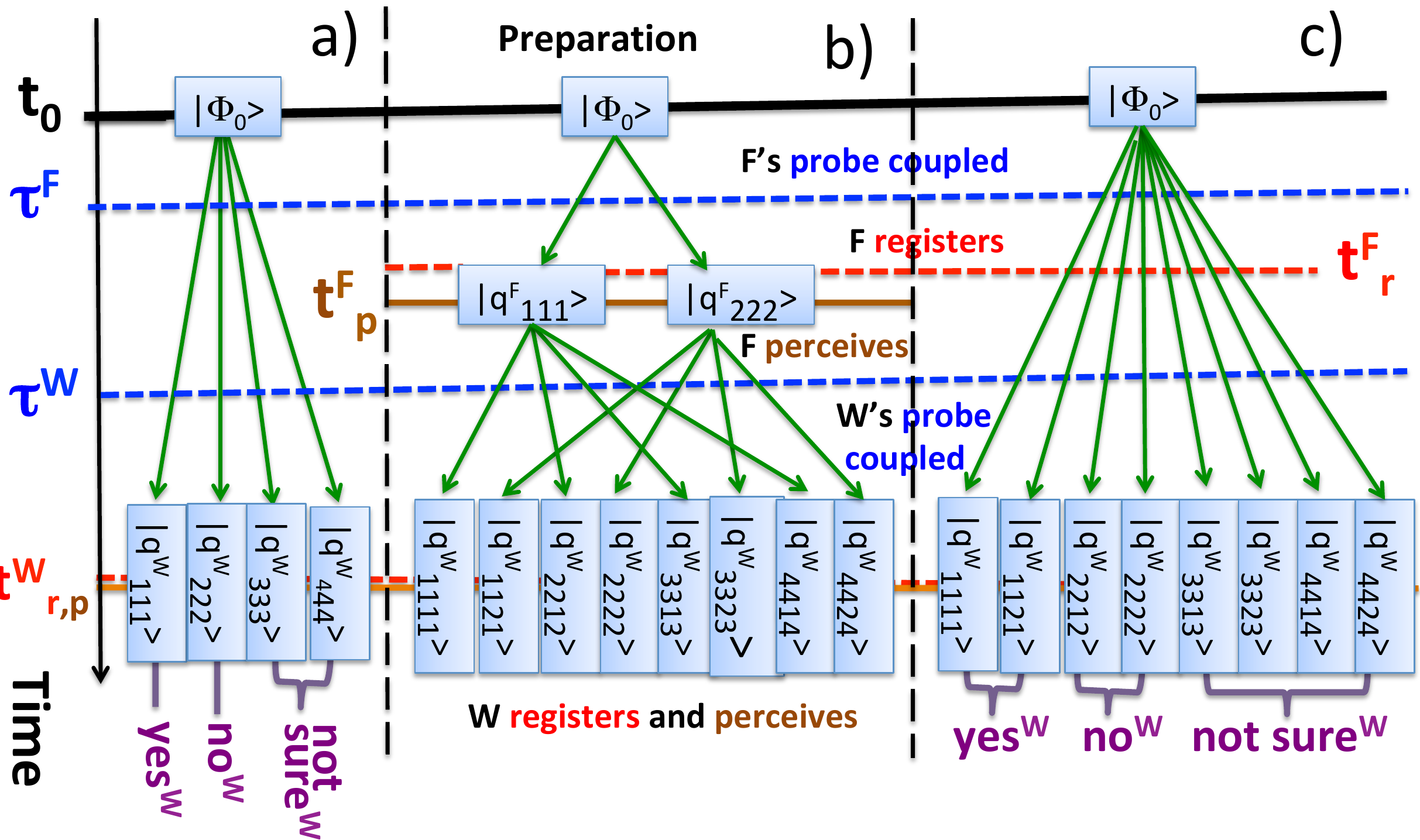}
\caption {Virtual paths in case a) F does not register, nor perceive his outcome, and W perceives his outcome;
b) both F and W register and perceive their respective outcomes; c) F only registers his outcome, and W registers and perceives his.
In the scenario a) W sees interference on his results. In b) and c) this interference is destroyed, since F's memory carries 
a record of his outcome, even if it has not been perceived. }
\label{fig:FIG1}
\end{figure}
%%%%%%%%%%%%%%%%%%%
\begin{eqnarray} \label{063}
\la q^W_{ij1}|\U(t^W_p,t_0)|\mu^W_0\ra|d^W_0\ra |d^F_0\ra|s_0\ra=\delta_{i1}\delta_{j1}[\alpha^*A^S_1+\beta^*A^S_4]\q\q\q\q\q\q\q\q\q\q\q\q\q\q\q\n
%\delta_{j1}\delta_{k1}\times\q\q\q\q\q\q\q\q\q\n
%\left [\alpha^*\la s_1^W|\u^S(\tau^W,\tau^F)|s^F_1\ra
%\la s_1^F|\u^S(\tau^F,t_0)|s_0\ra
%+ \beta^*\la s_2^W|\u^S(\tau^W,\tau^F)|s^F_2\ra
%\la s_2^F|\u^S(\tau^F,t_0)|s_0\ra \right ],\n
%%%%%%%%%
\la q^W_{ij2}|\U(t^W_p,t_0)|\mu^W_0\ra|d^W_0\ra |d^F_0\ra|s_0\ra=\delta_{i2}\delta_{j2}[\beta A^S_1-\alpha A^S_4],\q\q\q\q\q\q\q\q\q\q\q\q\q\q\q\n
%%%%%%%%%%
\la q^W_{ij3}|\U(t^W_p,t_0)|\mu^W_0\ra|d^W_0\ra |d^F_0\ra|s_0\ra=\delta_{i3}\delta_{j3}[\alpha^*A^S_2+\beta^*A^S_3],
\q\q\q\q\q\q\q\q\q\q\q\q\q\q\n
%\delta_{j3}\delta_{k3}\times\q\q\q\q\q\q\q\q\q\n
%\left [\alpha^*\la s_1^W|\u^S(\tau^W,\tau^F)|s^F_2\ra
%\%la s_2^F|\u^S(\tau^F,t_0)|s_0\ra +\beta^*\la s_2^W|\u^S(\tau^W,\tau^F)|s^F_1\ra \la s_1^F|\u^S(\tau^F,t_0)|s_0\ra
% \right ],\n
%%%%%%%%%%%
\la q^W_{ij4}|\U(t^W_p,t_0)|\mu^W_0\ra|d^W_0\ra |d^F_0\ra|s_0\ra=\delta_{i4}\delta_{j4}[\beta A^S_2-\alpha A^S_3],\q\q\q\q\q\q\q\q\q\q\q\q\q\q\q
\end{eqnarray}
where
\begin{eqnarray} \label{063a}
A_1^S\equiv \la s_1^W|\u^S(\tau^W,\tau^F)|s^F_1\ra\la s_1^F|\u^S(\tau^F,t_0)|s_0\ra, \q\q\q\q\q\q\q\q\q\q\q\q\q\q\q\n 
A_2^S\equiv \la s_1^W|\u^S(\tau^W,\tau^F)|s^F_2\ra \la s_2^F|\u^S(\tau^F,t_0)|s_0\ra,\q\q\q\q\q\q\q\q\q\q\q\q\q\q\q\n 
A_3^S\equiv\la s_2^W|\u^S(\tau^W,\tau^F)|s^F_1\ra\la s_1^F|\u^S(\tau^F,t_0)|s_0\ra,\q\q\q\q\q\q\q\q\q\q\q\q\q\q\q \n
A_4^S\equiv\la s_2^W|\u^S(\tau^W,\tau^F)|s^F_2\ra \la s_2^F|\u^S(\tau^F,t_0)|s_0\ra.\q\q\q\q\q\q\q\q\q\q\q\q\q\q\q
\end{eqnarray}
The four paths in the $64$-dimensional Hilbert space of  the composite {\it \{system+F's probe+W's probe+W's memory \}}  with non-zero amplitudes are shown in Fig.1a.
Then, by (\ref{04}), we have
\begin{eqnarray} \label{073}
P_1(yes^W)=|\alpha^*A^S_1+\beta^*A^S_4|^2= |\alpha|^2|A^S_1|^2+|\beta|^2|A^S_4|^2+2\R[\alpha^*\beta A^SA^{s*}_4],\q\q\n
P_1(no^W)=|\beta A^S_1-\alpha A^S_4|^2,= |\beta|^2|A^S_1|^2+|\alpha|^2|A^S_4|^2-2\R[\alpha^*\beta A^SA^{s*}_4]\q\q\q\n
P_1(\{not\q sure\}^W)= |\alpha^*A^S_2+\beta^*A^S_3|^2+|\beta A^S_2-\alpha A^3_4|^2.\q\q\q\q\q
\end{eqnarray}
These three probabilities add up to one, as they should, since $\sum_{i=1}^4|A_i^S|^2=1$, $|\alpha|^2+|\beta|^2=1$, and $\la s_0|s_0\ra=1$.
 \newline 
 {\it Scenario (B): Both F and W register and perceive their outcomes.} 
Next consider the case where both F and W become aware of their outcomes at the same time the outcomes become recorded in their memories,
\begin{eqnarray} \label{times}
t^F_p=t_r^F+\epsilon, \q \epsilon \to 0.
\end{eqnarray}
There are now two sets of possible outcomes and, according to Eqs.(\ref{01})-(\ref{04}), we need the matrix elements of two operators, 
$\u(t^F_p,t_0)$, which takes into account developments (\ref{013}) and (\ref{023}), and $\u(t^W_p,t^F_p)$, which includes 
(\ref{043}) and (\ref{053}). To describe F's relation with his memory, 
 we will use an operator  
\begin{eqnarray} \label{Fop}
\hat Q^F=|\mu^F_1\ra \la \mu^F_1|+2|\mu^F_2\ra \la \mu^F_2|, 
\end{eqnarray}
whose eigenvalues are interpreted as $Q^F_1=1\to yes^F$, $Q^F_2=2\to no^F$. Since W's probe and memory remain unchanged until $\tau^W> t^F_p$
we can choose eigenstates of  $\hat Q^F$ to be  
\begin{eqnarray} \label{Wst2}
|q^F_{ijk}\ra = |\mu^W_0\ra|\mu^F_k\ra|d^F_j\ra |d^W_0\ra|\underline  s_i^F\ra,\q i,j,k=1,2,
\end{eqnarray}
where $|\underline  s^F_i\ra \equiv \u^S(t^F_p,\tau^F)|s_i^F\ra$. 
To include F's memory (no longer idle)  in the calculation of W's probabilities, we choose the eigenstates of $\hat Q^W$ in Eq.(\ref{Wop})
in the form 
\begin{eqnarray} \label{Wst2}
|q^W_{ijkl}\ra=|\mu^W_l\ra|\mu^F_k\ra|d^W_j\ra|\underline {\phi}^W_i\ra, \q 
i,j,l=1,2,3,4, \q k=1,2.
\end{eqnarray}
With this, we have
\begin{eqnarray} \label{0103}
\la q^F_{ijk}|\u(t^F_p,t_0)|\mu^F_0\ra |d^F_0\ra|s_0\ra=\delta_{ij}\delta_{jk}\la s^F_i|\u^S(\tau^F,t_0)|s_0\ra,
\end{eqnarray}
\begin{eqnarray} \label{0113}
\la q^W_{i'j'k'l'} |\u(t^W_p,t^F_p)|q^F_{ijk}\ra=\delta_{i'j'}\delta_{j'l'}\delta_{k'k}\la \phi^W_{i'}|\u^S(\tau^W,\tau^F)|s^F_i\ra|d^F_i\ra,
\end{eqnarray}
so that there are eight final states $| q^W_{i'i'k'i'}\ra$, which can be reached from the initial $|\Phi_0\ra$ via 
eight paths shown in Fig.1b. In particular we have 
\begin{eqnarray} \label{0123}
P(yes^W,yes^F)=|\la q^W_{1111} |\u(t^W_p,t^F_p)|q^F_{111}\ra\la q^F_{111}|\u(t^F_p,t_0)|\Phi_0\ra|^2=|\alpha|^2|A_1^S|^2,\n
P(yes^W,no^F)=|\la q^W_{1121} |\u(t^W_p,t^F_p)|q^F_{222}\ra\la q^F_{222}|\u(t^F_p,t_0)|\Phi_0\ra|^2=|\beta|^2|A_4^S|^2, 
\end{eqnarray}
where the amplitudes $A^S_{1,2}$ are defined in Eqs.(\ref{063a}).  Thus, the net probability of W perceiving a result $yes^W$, 
\begin{eqnarray} \label{0133}
P_2(yes^W)=P(yes^W,yes^F)+P(yes^W,no^F) =|\alpha|^2|A_1^S|^2+|\beta|^2|A_4^S|^2, 
\end{eqnarray}
is different from the one in Eq.(\ref{073}), apparently changed by the fact that F had previously {\it perceived} his outcome. 
%%%%%%%%%%%%%%%%%%%%%%%%%%%%%%%
\newline
{\it Scenario (C): F only registers his outcome, and W registers and perceives his.}
We would have failed in our task of setting Observer's consciousness aside, if a mere act of F's perception could alter 
W's statistics. There is, however, no danger of that happening, as seen from the example where F's memory carries a record of his result,
but F is not supposed to perceive it before the experiment is finished.  
According to the rules of Sect.II, this case is formally different from the already discussed scenarios $(A)$ and $(B)$. 
Indeed, there is only 
one (W's) set of perceived outcomes ($yesW$, $no^W$, or $\{ not \q sure\}^W$), and eight paths connecting $|\Phi_0\ra$ with the final states
$|q^W_{iiki}\ra$, $i=1,2,3,4$, $k=1,2$, shown in Fig.1c. Now the probability of W perceiving an outcome $yes^W$ is 
\begin{eqnarray} \label{0143}
P_3(yes^W) =|\la q^W_{1111}|\u(t^W_p,t_0)|\Phi_0\ra|^2+|\la q^W_{1121}|\u(t^W_p,t_0)|\Phi_0\ra|^2=|\alpha|^2|A_1^S|^2+|\beta|^2|A_4^S|^2,
\end{eqnarray}
which is the same as $P_2(yes^W)$  in Eq.(\ref{0133}), but differs from $P_1(yes^W)$  in Eq.(\ref{073}) by the absence of the interfering term
$2\R\left[ \alpha \beta^*A^S_1A^{s*}_4\right]$. 
%%%%%%%%%%%%%%%%%%%%%%%%%
\section{Feynman's photon, future possibilities, destroyed records, and missed opportunities}
A brief summary is in order. 
A material record is carried by a system, to which the calculation of Sect.II ascribes a different orthogonal state, for each scenario, considered there.
The world \e{material} is chosen to emphasise that the recording system is a material object, and nothing essential for calculating the probabilities  is consigned to the Observer's consciousness, where quantum theory has no jurisdiction. A simple example of such a record was given by Feynman in \cite{FeynL} where, in a double-slit experiment with electrons, a photon would end up in distinguishable (orthogonal) states, depending on the slit at which it was scattered  by the passing particle. This alone will destroy the interference pattern on the screen { even if the photon is never detected} -
{\it \e{At the end of the process you may say that you 'don't want to look at the photon. That's your business, but you still do not add the amplitudes...}}\cite{FeynL}. 
\newline
The case of the previous Section are conceptually similar to the above example, with F's memory playing the role of Feynman's photon. 
Comparing the scenarios A and B, we  note that  quantum theory automatically  accounts for the effect of previous (F's) perceptions 
 on the final (W's) statistics {\it provided} the act o perception is accompanied by creation of a record 
in a material object representing Observers memory. As such, the memory is akin to any other object, carrying information 
about the observation's outcome. For example, an Observer, not wishing to rely only on his/her memory, could decide to leave an additional note, 
e.g., by preparing a spin up a given axis, if the result is a $yes$, or down that axis, if it is a $no$. In addition,  he/she may decide to communicate  
the outcome to a friend, whose memory will be changed accordingly upon receiving this information. In all these cases, quantum mechanics will need only to take into account 
the records' degrees of freedom, in order to be able to make the correct prediction using the rules of Sect.II. Moreover, as our scenario C shows, 
the actual act of perception is not necessary. Even if a record, accessible to an Observer {\it in principle} \cite{FeynL} in future, is created 
by an inanimate device, the final statistics will look as if the outcome of and intermediate measurement had been experienced. 
\newline
%In doing so, the theory draws a line between Observer's consciousness and his/her material memory, leaving the former
%firmly outside its scope. An Observer is by no means forbidden from falling in love, writing poetry, or praying to God - but what concerns
%his/her experiences of a material world must be  subject to quantum rules of Sect....  
Feynman's example may have interesting implications for an experiment where a macroscopic (F's) probe becomes entangled with a small quantum system, 
such as an atom, or a spin, and W attempts to erase the information by entangling his probe to both F's probe and the system, as in our scenario B.
There W  failed to do so, because of the persisting record in F's memory. But the same result would have been obtained even without 
 F present, provided a single photon had been scattered by F's probe, and then escaped W's manipulations.  Isolating a macroscopic device from 
all microscopic influences may be an extremely difficult task, even in the absence of a macroscopic environment, whose presence 
is often assumed in decoherence theories \cite{DC}.
\newline 
One can still ask what would happen if W manages to entangle the system, F's probe, F's memory, the memory of F's friend with whom 
F shared his outcome, and all the photons which were scattered during the experiment? At least, there is no formal contradiction. 
With all material records destroyed, the knowledge of what actually happened in the experiment will be irretrievably lost. 
This answer relies on an assumption we had to make, namely that no information about the physical world 
can be stored anywhere beyond the reach of the theory \cite{DSepl2}.
% e.g., in the Observer's consciousness . 
There is an agrrement with purely classical reasoning - in a classical world, with {\it all} witnesses dead, and {\it all} records destroyed it should be impossible to know what did actually happen. 
\newline
Finally, one may ask \e{did the information about F's outcome exist before W destroyed all the records?}.
We recall that in the example of the previous Section quantum mechanics provides probabilities for three hypothetical sequences of actual events. According to Feynman \cite{FeynL}, a  probability is desired for when the experiment is 'finished', i.e., when W 
perceives his outcome at $t=t^W_p$. In the scenario $(A)$, F had an opportunity to look at his probe and obtain his outcome at any 
$\tau^F < t < \tau^W$, but missed it. Were he to take this opportunity, we would be in the scenario $(B)$.
Finally, if a perceivable record were to be made without F's knowledge, the theory would consider 
scenario $(C)$, where there remains, at least 
{\it in principle} \cite{FeynL},  a possibility of learning about the outcome in future, at some $t>t^W_p$. 
%%%%%%%%%%%%%%%%%%%%%%%%%%%%
\section{The Wigner's Friend problem}
In a much cited paper \cite{Wig}, Wigner (W) considered the following situation. 
A quantum system $S$, initially is a state $\psi=\alpha |s_1\ra+ \beta |s_2\ra$, is coupled to a probe $D$, and becomes entangled with it. 
The state of the composite {\it S+D} becomes $\phi=\alpha |d_1\ra |s_1\ra+ \beta |d_2\ra|s_2\ra$, which is fully acceptable so far.
If the probe is replaced by a conscious Observer, Wigner's Friend (F), so that the states $|d_1\ra$ and $|d_2\ra$ correspond to him/her having seen the system in conditions $|s_1\ra$ and $|s_2\ra$, respectively, the situation changes.
Should Wigner ask his Friend about the condition of $S$ a when he saw it, F has to give one of the two possible answers. 
 The entangled state no longer makes sense, since it corresponds to neither $|s_1\ra$ , nor $|s_2\ra$, 
 and appears to imply that the {\it \e{friend was in a state of suspended animation}}\cite{Wig} until asked by Wigner. 
 The solution proposed in \cite{Wig} was to suggest that the unsuitable state $\phi$ be replaced  by a mixture  
 $|\alpha|^2 |d_1\ra |s_1\ra\la d_1|\la s_1|+|\beta|^2 |d_2\ra |s_2\ra\la d_2|\la s_2|$, which now describes two exclusive 
 alternatives available to F. This change, concludes Wigner, must be effected by Friend's consciousness exerting influence upon physical world, 
 hence the necessity to make quantum equations of motion {\it\e{grossly non-linear if conscious beings enter the picture}} \cite{Wig}.
 \newline
 Let us reconsider the situation in a slightly different, yet equivalent form, after making one additional assumption.
 %After preparation, Friend couples his visible probe $D$  to a quantum system $S$,
 % at $t=\tau^F$, 
 %registers the result and perceives the outcome all roughly at the same time at $t=t_1>\tau^F$, and lets the probe and the system alone until a $t_2 <t_1$, when Wigner asks him what he saw. 
%We will be able leave F's consciousness outside the picture, 
%except for one assumption.
 It is contrary to our experience that a person should be conscious of all information stored in his/her memory at all times. The sequence of events, about which W wants to make predictions, must therefore look like this. At $t=t_0$, F prepares the system and his probe,
and keeps them apart  
until  $t=t_1$.  At $t_1$, he couples his probe to the system, looks at the probe, and consigns the outcome to his memory $\M$. 
He then goes on thinking 
about unrelated matters, such as football or the state of the economy. After $t_1$,  $S$ and $D$ may  interact with each other, but not with $\M$. 
At a $t=t_2> t_1$, W asks F about what he saw at $t=t_1$, so F has to consult his memory again, before coming up with  an answer. 
\newline 
The situation is easily analysed by using the prescriptions of Sect.II.
 We should consider a composite {\it \{system+F's probe+F's memory\}}.
(If F were to make other records of his outcome, these would need to be included as well).
As before, F's perceived  outcomes will be $yes$ or $no$, F's memory will couple to the probe as  $|\mu_0\ra|d\ra \to \la d_1|d\ra|d_1\ra|\mu_1\ra+
\la d_2|d\ra|d_2 \ra|\mu_2\ra$, and the eigenvalues of F's operator $\hat Q =|\mu_1\ra\la \mu_1|+2|\mu_2\ra\la \mu_2|$ will be interpreted 
as $Q_1=1\to yes$ and  $Q_2=2\to no$. There are two perceived outcomes, one at $t=t_1$, when the result was first registered, 
and one at $t=t_2$, when Friend needs to answer Wigner's question. We require two sets of states for the composite,
\begin{eqnarray} \label{041w}
|q^{(1)}_{ijk}\ra=|\mu_k\ra|d_j\ra|s_i\ra, \q i,l,k,=1,2,
\end{eqnarray}
and
%\end{document}  
\begin{eqnarray} \label{042w}
|q^{(2)}_{ik}\ra=|\mu_k\ra|\varphi_{i}\ra, \q i,k=1,2,\q{\varphi}_{i}\equiv \u^{D+S}(t_2,t_1)|d_i\ra|s_i\ra.
\end{eqnarray}
In (\ref{041w}) we have assumed that F's probe and the system do not interact before $t_1$, and the evolution operator
$\u^{D+S}(t_2,t_1)$ in (\ref{041w}) accounts for any interaction between $S$ and $D$ that may occur after F becomes aware of  his outcome for the first time. 
For the amplitudes for the virtual paths $\{q^{(2)}_{i'k'} \gets q^{(1)}_{ijk} \gets \mu_0 d_0 s_0\}$ we have
\begin{eqnarray} \label{043w}
A(q^{(2)}_{i'k'} \gets q^{(1)}_{ijk} \gets \mu_0 d_0 s_0)= \la q^{(2)}_{i'k'}| \u(t_2,t_1)|q^{(1)}_{ijk}\ra \la q^{(1)}_{ijk}|\u(t_1,t_0)|\mu_0\ra|d_0\ra|s_0\ra\n
= \delta_{kk'}  \delta_{ii'}\delta_{jk}\delta_{ij}\la s_i|\u^S(t_1,t_0)|s_0\ra\q\q\q\q\q\q\q\q\q\q\q\q\q\q\q\q\q\q
\end{eqnarray}
There are two virtual paths with non-zero amplitudes, and the probabilities of F perceiving outcomes $yes^{1,2}/no^{1,2}$ at $t_{1,2}$ 
are
\begin{eqnarray} \label{044w}
P(yes^{2}, yes^{1}) = |A(q^{(2)}_{11} \gets q^{(1)}_{111} \gets \mu_0 d_0 s_0)|^2=|\la s_1|\u^S(t_1,t_0)|s_0\ra|^2,\n
P(yes^{2}, no^{1}) = |A(q^{(2)}_{12} \gets q^{(1)}_{222} \gets \mu_0 d_0 s_0)|^2=0,\q\q\q\q\q\q\q\q\n
P(no^{2}, yes^{1}) = |A(q^{(2)}_{21} \gets q^{(1)}_{111} \gets \mu_0 d_0 s_0)|^2=0,\q\q\q\q\q\q\q\q\n
P(no^{2}, no^{1}) = |A(q^{(2)}_{22} \gets q^{(1)}_{222} \gets \mu_0 d_0 s_0)|^2=|\la s_2|\u^S(t_1,t_0)|s_0\ra|^2,\q
\end{eqnarray}
which is the expected result. 
First,  Friend looks at his probe and consigns the outcome to his memory. When asked about it by Wigner, F consults the memory (or any other 
material record he may have produced) and gives an honest reply. Quantum mechanics duly takes notice of any records produced, and if 
the rules of Sect.II are accepted as its basic principle, no revision or extension of the existing formalism is required.
%%%%%%%%%%%%%%%%%%%%%%%%%%%
\section{ An interference  gedankenexperiment}
In a 1995 paper \cite{Deu}, Deutsch, very much in the spirit of Section III, studied the  "collapse" of the wave function, i.e., the process 
whereby a superposition $\sum c_i |\Phi_i\ra$ goes into a single term, say, $|\Phi_{i_0}\ra$, which corresponds to the 
actually observed value of the measured operator. In a slightly simplified form, the proposed experiment consists of coupling a system $S$  to a probe $D$, 
then measuring an operator, able to  detect  whether the coupling took place, storing the outcome, and then
reversing the system-probe evolution.
 In the scheme of \cite{Deu}, coupling of the probe is equivalent to 
\e{subsystem $D$ measuring subsystem $S$}, and if this measurement is "complete", the state of $\{ S+D\}$ would collapse  \cite{Deu}, 
and will not be restored to its initial form by the reversed evolution.
The  experiment is finished with measuring $S+D$, in a different basis,
 so that the statistics of this last measurement would indicate 
whether the composite $\{S+D\}$  ends up in a pure, or in a mixed state.
In particular, one might ask whether knowing that the first measurement took place, but not its outcome, would have and effect 
on the statistics of the last measurement. 
\newline
Unlike the author of \cite{Deu}, we are not interested in the virtues,  or otherwise, of the Copenhagen and Everett's many-world interpretations.
Our aim is to demonstrate that by applying the rules of Sect.II, we can get a definite answer without mentioning either of the two schools of thought directly.
As before, we will employ two Observers, F and W, the former to establish that the measurement coupling did take place, and the latter to collect the final statistics. We will require three probes, $D$, $D^F$ and $D^W$, a two-level system $S$, and assume that only $S$ has 
its own nontrivial dynamics. As before, at $t_0=0$ a composite {\it \{system+probe+F's probe+W's probe+F's memory+W's memory \}} is prepared in an initial state
\begin{eqnarray} \label{015}
|\Phi_0\ra =|\mu^W_0\ra|\mu^F_0\ra |d^W_0\ra |d^F_0\ra |d_0\ra|s_0\ra.
\end{eqnarray}
Then at $\tau > t_0$,   $S$ is entangled with the probe $D$, according to 
\begin{eqnarray} \label{025}
|d_0\ra|s\ra \to \la s_1|s\ra |d_1\ra|s_1\ra+\la s_2|s\ra |d_2\ra|s_2\ra, \q\q\q\q\q \n 
\la s_j|s_i\ra=\delta_{ij}, \q i=1,2,\q \la d _n|d_m\ra=\delta_{mn},\q m,n=0,1,2.
\end{eqnarray}
Later still, at $t_1> \tau$, F's probe is entangled with the composite $\{ S+D\}$, and F immediately registers and perceives his outcome.
This development is described as
\begin{eqnarray} \label{035}
|\mu_0^F\ra|d_0^F\ra|\phi\ra \to |\mu_1^F\ra|d_1^F\ra\left [\alpha_{11} |d_1\ra|s_1\ra+\alpha_{22} |d_2\ra|s_2\ra\right]+\n
|\mu_2^F\ra|d_2^F\ra\left [\sum_{i=1}^2\sum_{j=0}^2 (1-\delta_{j1}\delta_{i1})(1-\delta_{j2}\delta_{i2})\alpha_{ij}|d_j\ra |s_i\ra\right],
\end{eqnarray}
where $|\phi\ra$ is an arbitrary state  of the $\{S+D\}$, and $\alpha_{ij} \equiv \la d_j|\la s_i|\phi\ra$, $i=1,2$, $j=0,1,2$. 
Equation (\ref{035}) makes F a reliable witness  - his memory is in a state 
$|\mu_1^F\ra$ if the coupling  (\ref{025}) was applied (coupling $on$), and in a distinguishable state $|\mu_2^F\ra$, if it was not (coupling $off$). 
After  $t_1$, the evolution of $S$ and $D$, up to the moment F couples his probe, is reversed.
Until $t=t_1+(t_1-\tau)$ the system's evolution operator is
$\u^S(2t_1-\tau,t_1)=[\u^{S}]^{-1}(t_1,\tau)$. At $t=2t_1-\tau$ the coupling (\ref{025}) is reversed, to wit
\begin{eqnarray} \label{045}
|d_i\ra|s_i\ra\to |d_0\ra|s_i\ra, \q i=1,2.
\end{eqnarray}
 From $t=2t_1-\tau$ to $t_2= 2t_1$, at which W perceives his outcome, we have $\u^S(t_2, 2t_1-\tau)=[\u^S]^{-1}(\tau,0)$.
%\newline
At $t_2=2t_1$, W may decide to explore the odds on finding the $\{S+ D\}$ in the initial  state $|d_0\ra|s_0\ra$, $P(\text{back  to }|d_0\ra| s_0\ra)$,
by coupling and registering his probe according to
\begin{eqnarray} \label{045a}
|\mu_0^W\ra|d^W_0\ra|\phi^{S+D}\ra \to \la s_0|\la d_0||\phi^{S+D}\ra\times |\mu_1^W\ra|d^W_1\ra|s_0\ra |d_0\ra+...,
%\la s_j|s_i\ra=\delta_{ij}, \q i=1,2,\q \la d _n|d_m\ra=\delta_{mn},\q m,n=0,1,2.
\end{eqnarray}
where $|\phi^{S+D}\ra$ is an arbitrary state of the $\{S+ D\}$, and we omitted the terms, containing the five remaining orthogonal states 
of the composite. 
%, with an $|sF\ra \ne |s_0\ra$. 
If the entanglement of probe $D$ with $S$ did not constitute a \e{complete measurement} [from (\ref{035}) we know that the interaction 
between $S$ and $D$  did take place], the composite will be restored to  $|d_0\ra|s_0\ra$ with a probability 
 \begin{eqnarray} \label{055}
P_1(\text{back  to }|d_0\ra| s_0\ra)=
%\q\q\q\q
%\q\q\q\q\q\q\q\q\q\q\q\q\q\q\q\q\q\q\q\q\q\q\q\n
%|\la s^0| [\u^{s}(\tau,0)]^{-1}|s_1\ra
\left [|\la s_1|\u^{S}(\tau,0)|s_0 \ra|^2 +
%\la s^W| [\u^{s}(\tau,0)]^{-1}|s_2\ra
|\la s_2|\u^{S}(\tau,0)|s_0 \ra|^2\right ]^2= \la s_0|s_0\ra^2=1.
%|\la s^W|s_0\ra|^2
% \n
%(|\la s_1|\u^S(\tau)|s_0\ra|^2+|\la s_2|\u^S(\tau)|s_0\ra|^2)^2=1.\q\q\q\q\q\q\q\q\q\q\q\q\q\q\q
\end{eqnarray}
If a complete measurement takes place, at $t=t_1$, the \e{collapsed state} of the composite will have to be
either $\u^S(t_1,\tau)|s_1\ra|d_1\ra$, with a probability $|\la s_1|\u^S(\tau)|s_0\ra|^2$,  or $\u^S(t_1,\tau)|s_2\ra|d_2\ra$, with a probability  $|\la s_2|\u^S(\tau)|s_0\ra|^2$, and W's odds,  
 \begin{eqnarray} \label{056}
P_2(\text{back  to }|d_0\ra| s_0\ra)=
%\q\q\q\q\q\q\q\q\q\q\q\q\q\q\q\q\q\q\q\q\q\n
|\la s_1|\u^S(\tau,0)|s_0\ra|^4+|\la s_2|\u^S(\tau,0)|s_0\ra|^4,%^2=1.\q\q\q\q\q\q\q\q\q\q\q\q\q\q\q
\end{eqnarray}
will differ from $P_1$ in (\ref{055})  by an interference term $P_1(\text{back  to }|d_0\ra| s_0\ra)-P_2(\text{back  to }|d_0\ra| s_0\ra)=2|\la s_1|\u^S(\tau)|s_0\ra|^2|\la s_1|\u^S(\tau)|s_0\ra|^2$. Thus, the question is whether Eq. (\ref{055}) or  Eq. (\ref{056}) will yield the correct answer, given that we {\it know}
that the \e{subsystem $D$ has measured the subsystem $S$ at $t=\tau$,} but do not know the measurement's outcome?
\newline 
The answer, easily found by applying the rules of Sect.II, is shorter than the question it took us some time to formulate. Figure 2 shows two virtual paths connecting the initial and the final states with non-zero  the amplitudes  ($i=1,2$)
%, $t_{1,2}=t_{1,2}+\epsilon$, $\epsilon \to 0$ )
  \begin{eqnarray} \label{057}
A_i\equiv \la\mu_{1}^W| \la\mu_{1}^F|\la d_{1}^W| \la d_{1}^F| \la d_{0}|\la s_0|\u(t_2,t_1)|\mu_{0}^W\ra|\mu_{1}^F\ra|d_{0}^W\ra |d_{1}^F \ra |d_{i}\ra |s_i\ra
\times\q\q\q\q\q\q\q\q\q\q\q\n 
\la\mu_{0}^W| \la\mu_{1}^F|\la d_{0}^W| \la d_{1}^F| \la d_{i}|\la s_i|\u(t_1,0)|\mu_{0}^W\ra|\mu_{0}^F\ra|d_{0}^W\ra |d_{0}^F \ra |d_{0}\ra |s_0\ra=
%\n
\la s_0| [\u^{S}(\tau,0)]^{-1}|s_i\ra\la s_i| \u^{S}(\tau,0)|s_0\ra
\end{eqnarray}
%%%%%%%%%%%%%%%%%%%
\begin{figure}[h]
\includegraphics[angle=0,width=12cm, height= 6cm]{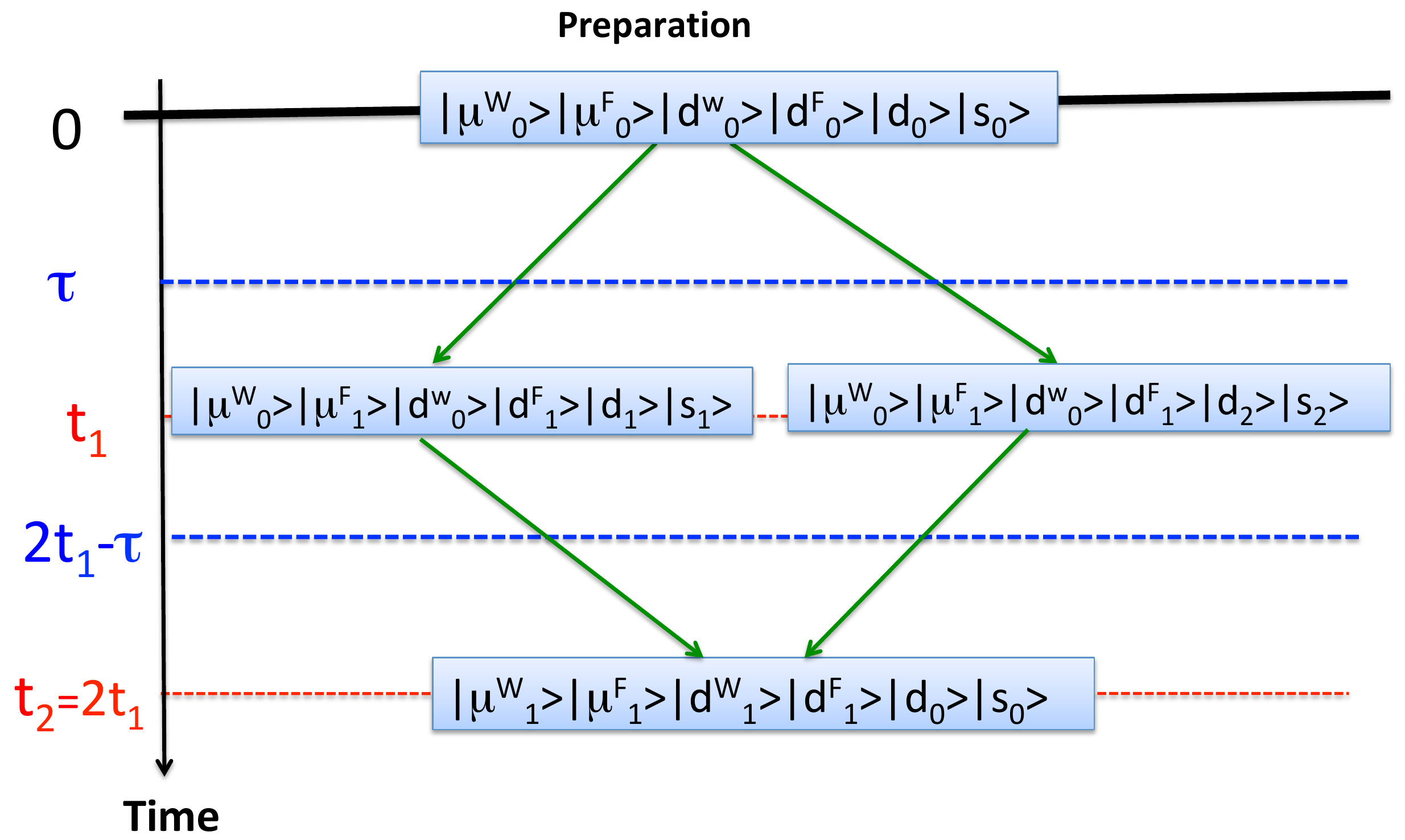}
\caption {Two virtual paths in the interference experiment of Sect.VIII. 
%$d_j\ra$ correspond to the subsystem $D$, 
$|d_j^{F,W}\ra$ and $|\mu_k^{F,W}\ra$, $j,k=0,1,2$ are the states of F's and W's the probes and memories, respectively. }
\label{fig:FIG1}
\end{figure}
%%%%%%%%%%%%%%%%%%% 
F's probe probe does not distinguish between the two scenarios shown in Fig. 3, and neither can F, 
%whose relation with the memory is
%described. e.g.,  by an operator $\hat Q^F= |\mu^F_1\ra\la\mu^F_1|+2|\mu^F_2\ra\la\mu^F_2|$, 
%$Q^F_1=1\to {"on"}$, $Q^F_2=2\to {"off"}$,
 who may only know if the \e{measurement of S by D} did take place. 
%Summing only the degeneracies of $Q^F_1$, 
Adding the amplitudes (\ref{057})
we recover the correct result (\ref{055}).
\begin{eqnarray} \label{059}
P(\text{back  to }|d_0\ra| s_0\ra)=|A_1+A_2|^2 =P_1(back \q to \q |d_0\ra| s_0\ra).
\end{eqnarray}
This is the case of the Feynman's photon, discussed in Sect.VI, with a difference that now the photon is always scattered into the same
state, regardless of the slit chosen by the electron.  Finding a scattered photon will signal the presence of a passing electron, but since no record of the path taken will be produced, an interfering pattern will be seen on the screen. 
%%%%%%%%%%%%%%%%%%%%%%%%%%%%%
\section{Reduction to the Hilbert space of the smallest system. The Von Neumann's boundary}
In classical mechanics an observation is expected to yield information about the observed system \e{on its own}, 
i.e., not influenced by being observed. A vestige of this principle in the quantum case is evident, e.g.,  from Eqs. (\ref{073}), (\ref{044w}), or (\ref{057}), 
where Observer's probabilities are expressed in terms of the probability amplitudes $A^S_i$,  referred to the system,
uncoupled from the probes used to measure it. However, this is as far as the analogy goes. The amplitudes are combined differently, depending on whether 
the probes have been involved or not.  This points towards a possibility of describing the measurements in a manner  more economical than
the one used up to now, namely by manipulating the system's amplitudes $A^S_i$, without referring to the probes' and the memory' degrees of freedom, which occupy so much space, for example, in Fig.1.
\newline
The idea is by no means new. In \cite{vN} von Neumann pointed out that quantum theory can successfully avoid analysing an Observer, 
provided the boundary between the Observer and the observed system can be displaced arbitrarily.
For example, it can be placed between Observer and his/her memory, $\M$, between the memory and the probe, $D$, or between 
the probe and the system $S$. The last option allows one to establish a direct correspondence between an Observer's experience, 
and a particular property, which Observer's theory ascribes to the system. Next we will apply this to a sequence of more than two measurements, 
$L>2$ relying, as before, on the rules of Sect.II. 
\newline
For an example, consider the case of Sect.V with a special choice $\alpha=1$ and $\beta=0$, so that the four states in Eq.(\ref{033})
 take the form
\begin{eqnarray} \label{061}
|\phi^W_1\ra= |s^W_1\ra|d^F_1\ra, \q |\phi^W_2\ra=- |s^W_2\ra|d^F_2\ra,\q
|\phi^W_3\ra=|s^W_1\ra|d^F_2\ra, \q |\phi^W_4\ra=- |s^W_2\ra|d^F_1\ra,
\end{eqnarray}
and F's probe, no longer affected by W's measurement, continues  to carry a record of the system's condition as it was at $t=\tau^F$. 
The degrees of freedom describing F's and W's memories, and W's probe, serve only to produce the Kronecker deltas in Eqs.(\ref{063}) and (\ref{0103}), 
and are readily taken into account by considering the paths in the Hilbert space of a smaller composite {\it \{system+F's probe\}}, shown in Fig.3a.
The paths end in different orthogonal final states $|\phi^W_j\ra$, $j=1,2,3,4$. According to the rules of Sect. II such paths cannot interfere, and can be endowed with probabilities, which are now the same for all three scenarios of Sect.V [cf. Eq.(\ref{03})], 
\begin{eqnarray} \label{062}
p(s^W_1d^F_1 \gets yes^F \gets d_0^F s_0)= |A^S_1|^2=|\la s^W_1|\u^S(\tau^W, \tau^F)|s^F_1\ra \la s^F_1|\u^S(\tau^F, t_0)|s_0\ra|^2,\n
p(s^W_1d^F_2 \gets no^F \gets d_0^F s_0)= |A^S_2|^2=|\la s^W_1|\u^S(\tau^W, \tau^F)|s^F_2\ra \la s^F_2|\u^S(\tau^F, t_0)|s_0\ra|^2,\n
p(s^W_2d^F_1 \gets yes^F \gets d_0^F s_0)= |A^S_3|^2=|\la s^W_2|\u^S(\tau^W, \tau^F)|s^F_1\ra \la s^F_1|\u^S(\tau^F, t_0)|s_0\ra|^2,\n
p(s^W_2d^F_2\gets no^F \gets d_0^F s_0)= |A^S_4|^2=|\la s^W_2|\u^S(\tau^W, \tau^F)|s^F_2\ra \la s^F_2|\u^S(\tau^F, t_0)|s_0\ra|^2.
\end{eqnarray}
%%%%%%%%%%%%%%%%%%%
\begin{figure}[h]
\includegraphics[angle=0,width=16cm, height= 9cm]{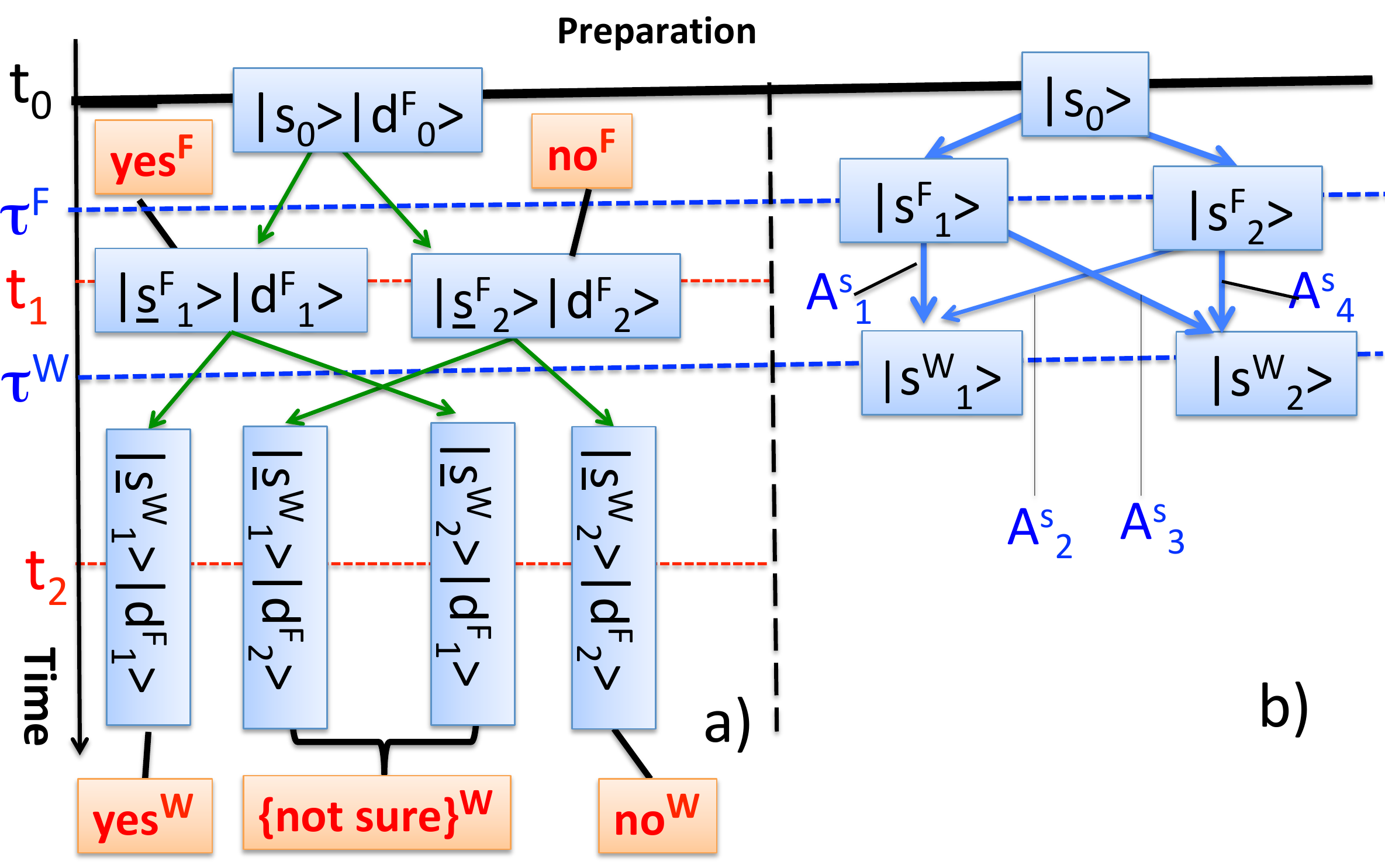}
\caption { a) Four real paths in the Hilbert space of {\it \{ system+F's\q probe\}}; b) four virtual paths in the Hilbert space of the 
system only. Coupling to F's probe does not change the values of the paths amplitudes $A_i^S$, but redirects the paths passing 
through through $|s^F_1\ra$ and $|s^F_2\ra$ to different final states in the larger Hilbert space, thus turning them into exclusive alternatives.}  
\label{fig:FIG1}
\end{figure}
%%%%%%%%%%%%%%%%%%%
It is readily seen that the degrees of freedom describing F's probe are also redundant since the r.h.s of Eqs.(\ref{062})
contains only the references to the system which makes transitions between the states $|s_0\ra$, $|s^F_i\ra$, and $|s^W_j\ra$.
Thus, to calculate the probabilities (\ref{062}) we may as well use a simpler diagram shown in Fig.3b. The diagram shows all 
virtual paths connecting $|s_0\ra$ with  $|s^W_j\ra$ at $t=\tau^W$, and passing through $|s^F_i\ra$ at $t=\tau^F$, when F's coupling was applied.
The only consequence of F's probe being involved is that now the paths leading to the same final states $|s^F_i\ra$ no longer interfere.
%of a double-slit experiment, where scattering of a photon (in our case, F's probe) allows to determine the manner is which the electron arrived at its final location \cite{FeynL}.
\newline 
This amounts to a general rule. One can apply the prescription of Sect.II to the observed system, Observers' probes, and their memories.
However, the same probabilities can be obtained by applying the same prescription directly to the system, represented by virtual paths in its (smaller) Hilbert space. In this case, the \e{behind the scenes} presence of the Observers and their probes is translated into 
destruction of interference between otherwise indistinguishable system's  routes. 
This is, of course, the Feynman's general rule for ascribing probabilities to distinguishable scenarios \cite{FeynL}.
\newline
Observers' memories, and their probes form  the so-called  von Neumann's chains \cite{vN}. They are distinguished 
 by a special type of interaction (\ref{A7}), 
coupling them to all other degrees of freedom, which together constitute the observed \e{system} (cf. Sect.  V).
\newline 
We note next that Eqs.(\ref{062}) contain no mention of the times $t_1$ and $t_2$, at which F an W perceive their respective results, and refer instead to the times $\tau^F$ and $\tau^W$, at which F's and W's respective probes were coupled to the system.
In the next Section we will discuss this lack of reference in more detail.
%%%%%%%%%%%%%%%%%%%%%%%%%
\section{Unitary evolution and the \e{in principle} principle}
In the example of the previous Section, F may well  decide not to register his probe before $t=t_2$, i.e., before the experiment is finished. 
This case is similar to the scenario $C$ of Sect.V in that there is only one perceived result, t(W's) which needs to be taken into account 
when applying the rules of Sect.II. In general, we can consider $L-1$ observers, all coupling their probes, $D^\l$, to a system $S$, at $t=\t_\l$, $\l=1,...L-1$, according to [cf. Eq.(\ref{A7})]
\begin{eqnarray} \label{071h}
|d^\l_0\ra |s\ra \to \sum_{m_\l=1} ^{M_\l} |d^\l_{m_\l}\ra \hat \pi_{m_\l}|s\ra. 
\end{eqnarray}
but failing to register their conditions before $t=t_L$, when the last, $L$-th,  Observer perceives his/her outcome. {In principle,} they could do it 
in the future, or maybe not do it at all.
\newline
Let the $L$-th observer measure a system's quantity 
$\mathcal S^L$, represented by an operator  $\hat S^L=|s_\nu\ra\la s_\nu|$ whose two eigenvalues are interpreted as $\S^L_1=1 \to yes^L$, 
 $\S^L_2=0 \to no^L$. The degrees of freedom, which describe the probe and the memory of the $L$-th Observer, can be left out of the calculation, as was discussed in the previous Section. For the probability of the $L$-th outcome \e{$yes$} we, therefore, have
\begin{eqnarray} \label{072h}
P(yes^L)= \sum_{m_1,...m_{L-1}=0} ^{M_1,...M_{L-1}}\left |\la s_\nu|\prod_{\l=1}^{L-1}
\la d^{\l}_{m_\l}|\u(t_L,t_0)|\Phi_0\ra\right |^2, \q |\Phi_0\ra \equiv |s_0\ra |\prod_{\l'=1}^{L-1}
| d^{\l'}_{0}\ra,
\end{eqnarray}
where the unitary evolution operator $\u(t_L,t_0)$ accounts for the couplings (\ref{071h}), as well as for the free evolution of the system $S$ 
between $\t_\l$ and $\t_{\l-1}$. The action of $\u(t_L,t_0)$ is fairly simple. It decomposes a free system's state $\u^S(t_L,t_0)|s_0\ra$ into
$M_1\times M_2...\times M_{L-1}$ in general 
non-orthogonal substates, 
\begin{eqnarray} \label{073h}
\u^S(t_L,t_0)|s_0\ra=\sum_{m_1,...m_{L-1}=1} ^{M_1,...M_{L-1}}|s_0(m_1,...m_{L-1})\ra, \q\q\q\q\n
|s_0(m_1,...m_{L-1})\ra \equiv \u^S(t_L,\t_{L-1})\prod_{\l=1}^{L-1}\hat \pi_{m_\l}\u^S(\t_\l,\t_{\l-1})|s_0\ra
\end{eqnarray}
and then \e{tags} each substate by multiplying it by one of the orthogonal probes' state $|d^1_{m_1}\ra ...|d^{L-1}_{m_{L-1}}\ra$, so that  we have
\begin{eqnarray} \label{074h}
|\Phi(t_2)\ra \equiv \u(t_L,t_0)|\Phi_0\ra = \u^S(t_L,t_{L-1})\sum_{m_1,...m_{L-1}=1} ^{M_1,...M_{L-1}}\left \{\prod_{\l=1}^{L-1} |d^{\l}_{m_{l}}	\ra\right \}
|s_0(m_1,...m_{L-1})\ra.
%\prod_{\l=1}^{L-1}\hat \pi_{m'_\l}\u^S(t_\l,t_{\l-1})|s_0\ra,
\end{eqnarray}
Now we can evaluate the matrix elements in Eq.(\ref{072h}), or adopt the view of Sect.III, and calculate $P(yes^L)$ 
%in Eq.(\ref{072h})
using the state $|\Phi(t_2)\ra$, obtained by a unitary evolution of $|\Phi_0\ra$, 
\begin{eqnarray} \label{075h}
P(yes^L)=
\text {tr}_{probes}
\left [ |s_\nu\ra \la s_\nu| |\Phi(t_2)\ra \la \Phi(t_2)| \right ]=\sum_{m_1,...m_{L-1}=1} ^{M_1,...M_{L-1}}|\la s_\nu|s_0(m_1,...m_{L-1})\ra|^2.
\end{eqnarray}
Reversing the last argument of the previous Section, we note that if the remaining $L-1$ Observers, each measuring the system's operators 
$\hat S^\l =\sum_{m_\l=1}^{M_\l}\S^\l_{m_\l}\hat \pi_{m_\l}$, $\l=1,...L-1$, {\it did} register there probes and perceived the outcomes in the course 
of the experiment, i.e., at $\t_\l < t_\l < t_L$, the probability would still be given by Eq.(\ref{075h}), 
\begin{eqnarray} \label{076h}
P(yes^L, \S^{L-1}_{m_{L-1}}... \S^{1}_{m_{1}})=|\la s_\nu|s_0(m_1,...m_{L-1})\ra|^2=|A^S(s_\nu ....\gets \S^{\l}_{m_{\l}}...\gets s_0)|^2 .
\end{eqnarray}
To arrive at  Eq.(\ref{076h}) we have, in fact,  used the Feynman's principle \cite{FeynL}: {\it \e{If you could}} \text{in principle}, {\it \e{distinguish 
the alternative final states (even though you don' bother to do so), the total, final probability is obtained by calculating the} 
\text{probability} {\it for each state (not the amplitude) and then adding them together.}} We applied it to the system's 
final states, $|s_0(m_1,...m_{L-1})\ra$, made distinguishable by the records, carried by the probes beyond 
the duration of the experiment, or, if one prefers,  
by the \e{tagging} evident  in Eq.(\ref{074h}). 
This saved us an effort 
of evaluating  a large number of (mostly trivial) matrix elements connecting the states in the (large) Hilbert space of the 
composite {\it \{system+all probes+all memories\}}, as would be necessary if the rules of Sect.II were to be applied 
%to account of the $L-1$ perceived results 
directly. 
\newline
Thus, replacing the effect produced by the $L-1$ intermediate Observers by an uninterrupted unitary evolution (\ref{074h})
helps simplify the calculation. 
It may also please the reader, to whom the \e{collapse} of a wave function a nuisance, and a potential problem.
His/her satisfaction would not, however, be complete. The need for the  last, the $L$-th, Observer to become aware of his/her outcome
implies projecting the overall state $|\Phi(t_2)\ra$ onto an orthogonal basis $\left (\prod_{\l=1}^{L-1} |d^{\l}_{m_{l}} \ra\right )|s_0(m_1,...m_{L-1})\ra$, 
and there is nothing we can do about it. Quantum rules of Sect.II  serve to predict statistical correlations between {\it at least two}
Observers' experiences (one of them disguised as \e{preparation}), and cannot be reduced further \cite{DSepl1}.
The content of these rules can, however, provide some insight into the matter. Calculation of matrix elements of  operators between states  in abstract Hilbert spaces
(which is all that is required)
does not rely on the concept of a constantly evolving \e{state}. 
Having a mental picture of such a state, and worrying about its fate after the $L$-th observer completes his/her observation, 
is just not necessary, if not futile,
like wondering about what {\it actually} happens to the fictional personage of a novel, once the last page is turned. 
%%%%%%%%%%%%%%%%%%%%%%%%%%%%%%%%
\section{ Where we  agree and disagree with the consistent histories approach}
Another method which uses the sequences of projectors similar to those in Eq.(\ref{011}) is the consistent histories approach (CHA) (see \cite{CHA}), and the Refs. therein),
and we will briefly discuss it here. At first glance, the CHA could  not be more different from our narrative. Indeed, while 
we aim at defining the probabilities of Observer's perceptions, i.e., of \e{certain (subjective) observations} \cite{vN}, the CHA, like \cite{Deu}, seeks a
\e{framework for reasoning about the properties of closed physical system} \cite{CHA1}, and gives no special role to a conscious Observer. 
According to the CHA, the probabilities $P(Q^{L}_{m_{L}}...\gets Q^{\l}_{m_{\l}}....\gets Q^1_{m_1}) $ can be ascribed to a sequence projectors (\ref{013}) (with $L-1$ changed to $L$), provided all partial evolutions
of an initial state $|q_0\ra$  result in mutually orthogonal states, 
\begin{eqnarray} \label{081}
\la q(Q^{L}_{m'_{L}}...\gets Q^{\l}_{m'_{\l}}....\gets Q^1_{m'_1})|q(Q^{L}_{m_{L}}...\gets Q^{\l}_{m_{\l}}....\gets Q^1_{m_1})\ra\n
= \delta_{m_1m'_1}...\delta_{m_\l m'_\l}...\delta_{m_{L} m'_{L}}P_{CHA}(Q^{L-1}_{m_{L}}...\gets Q^{\l}_{m_{\l}}....\gets Q^1_{m_1}), 
\end{eqnarray}
where [cf. Eq.(\ref{011})]
\begin{eqnarray} \label{082}
|q(Q^{L}_{m_{L}}...\gets Q^{\l}_{m_{\l}}....\gets Q^1_{m_1})\ra\equiv \u(Q^{L}_{m_{L}}...\gets Q^{\l}_{m_{\l}}....\gets Q^1_{m_1})|q_0\ra. 
\end{eqnarray}
Here we are not interested in the current discussion about the general merits of the CHA \cite{CHA2}, \cite{CHA3}, 
and will comment only on the significance of Eqs.(\ref{081}) and (\ref{082}) for our own discourse.
 
The projectors $ \hat \Pi^\l_{m_\l}$ in Eq.(\ref{011}) can stand for various things, and next we explore some of the possibilities.
\newline
{\it (i) The closed closed system, to which Eq.(\ref{081}) refers,  contains $L$ Observers, with their probes and memories.}
If so, we have
\begin{eqnarray} \label{083}
\hat \Pi^\l_{m_\l}=|\mu^\l_{m_\l}\ra \la \mu^\l_{m_\l}|\otimes|d^\l_{m_\l}\ra \la d^\l_{m_\l}|\otimes\hat \pi^s_{m_\l}, 
\end{eqnarray}
where the projector $\hat \pi^s_{m_\l}$ refers to the system only, and the $\u(t_\l,t_{\l-1})$ in Eq.(\ref{011}) accounts for the system's free evolution, 
 the coupling of the $\l$-th probe according to 
\begin{eqnarray} \label{084}
|d^\l_0\ra |s\ra \to 
\sum_{m_{\l}=1} ^{M_\l} |d_{m_\l}\ra \hat \pi_{m_\l}|s\ra, 
\end{eqnarray}
and for a similar coupling between the probe and the memory [cf. Eq.(\ref{023})].
In this case, the \e{consistency conditions} (\ref{081}) are satisfied, and $P_{CHA}(Q^{L}_{m_{L}}...\gets Q^{\l}_{m_{\l}}....\gets Q^1_{m_1})$ is just the probability (\ref{04}) for the  $L$ Observers to perceive their respective outcomes, at $t=t_\l$.
%regardless of what the last, [the $(L+1)$-th] outcome might be.
\newline 
{\it (ii) The closed closed system includes  the $L$-th Observer, and $L-1$ probes, not registered before the  experiment ends at $t=t_L$.}
This implies $\hat \Pi^\l_{m_\l}=|d^\l_{m_\l}\ra \la d^\l_{m_\l}|\otimes\hat \pi^s_{m_\l}$, for $\l=1,...L-1$, and 
(\ref{083}) for $\l=L$. For us the probabilities in Eq.(\ref{081}), identical to those in Eq.(\ref{075h}), have no individual meaning, 
since the $L-1$ intermediate outcomes were not perceived. 
However, their sum yields the correct odds on the $L$-th observer perceiving an outcome $Q^L_{m_L}$, 
%corresponding to $\S^L_{m_\l}$, 
\begin{eqnarray} \label{085}
P(\Q^L_{m_L})= \sum_{m_1,...m_{L-1}=1} ^{M_1,...M_{L-1}}P_{CHA}(Q^{L}_{m_{L}}...\gets Q^{\l}_{m_{\l}}....\gets Q^1_{m_1}).
\end{eqnarray}
We can also speculate about what would happen  if the remaining $L-1$ observers were to register their (protected) probes and perceive the outcomes
after the experiment is finished, i.e. at $\tau_\l > t_L$.
%\left |\la s_\nu|\prod_{\l=1}^{L-1}
%\la d^{\l}_{m_\l}|\u(t_L,t_0)|\Phi_0\ra\right |^2, \q |\Phi_0\ra \equiv |s_0\ra |\prod_{\l'=1}^{L-1}
%| d^{\l'}_{0}\ra,
\newline
 {\it (iii) The closed system includes only the system $S$ and the consistency conditions (\ref{081}) are satisfied.}
 Now we have $\hat \Pi^\l_{m_\l}=\hat \pi^s_{m_\l}$ and, since no Observers are present, have little to say about the 
 probabilities $P_{CHA}(\S^{L}_{m_{L}}...\gets \S^{\l}_{m_{\l}}....\gets \S^1_{m_1})$. We can, however, 
speculate about what would happen if the Observers were to join in. With only the $L$-th Observer present, his/her 
odds on seeing an outcome corresponding to $\S^{L}_{m_{L}}$  would be 
\begin{eqnarray} \label{085}
P(Q^L_{m_L})= \sum_{m_1,...m_{L-1}=1} ^{M_1,...M_{L-1}}P_{CHA}(\S^{L}_{m_{L}}...\gets \S^{\l}_{m_{\l}}....\gets \S^1_{m_1}).
\end{eqnarray}
If the remaining $L-1$ Observers were to join in, $\hat \Pi^\l_{m_\l} \to |\mu^\l_{m_\l}\ra \la \mu^\l_{m_\l}|\otimes|d^\l_{m_\l}\ra \la d^\l_{m_\l}|\otimes\hat \pi^s_{m_\l}$, the probabilities of their outcomes would be 
 \begin{eqnarray} \label{085}
P(Q^{L}_{m_{L}}...\gets Q^{\l}_{m_{\l}}....\gets Q^1_{m_1})= 
%\sum_{m_1,...m_{L-1}=1} ^{M_1,...M_{L-1}}
P_{CHA}(\S^{L}_{m_{L}}...\gets \S^{\l}_{m_{\l}}....\gets \S^1_{m_1}).
\end{eqnarray}
This is a \e{classical statistical} behaviour \cite{LEGG} - intermediate observations do not affect the final statistics, and the 
\e{which way?} question has an answer. As far as we are concerned, the consistency condition satisfied by the observed 
system, only indicates that no interference would be destroyed by observations of a particular type (but not by observations of {\it any} type), and some vestige of a classical behaviour can be retained. 
\newline 
{\it (iv) The closed system includes only the system $S$ and the consistency conditions (\ref{081}) are not satisfied.}
This is, probably, where our disagreement with the CHA is most evident. In itself, the failure to satisfy the condition (\ref{081}) means 
little to us, since we avoid to make statements about unobserved systems. 
We could, however, bring in all $L$ Observers, which would return us to the case (i). In the enlarged Hilbert space of 
the composite, the consistency condition would be satisfied, since the non-orthogonal system's substates $|q(\S^{L}_{m_{L}}...\gets \S^{\l}_{m_{\l}}....\gets \S^1_{m_1})\ra$ will acquire orthogonality upon multiplication by $|\mu^\l_{m_\l}\ra \otimes|d^\l_{m_\l}\ra$. 
Bringing in only the $L$-th observer will yield a different result, with his/her distribution $P_{\text{ one}}(Q^L_{m_L})$ 
not beeing a marginal of $P_{\text{all}}(Q^{L}_{m_{L}}...\gets Q^{\l}_{m_{\l}}....\gets Q^1_{m_1})$.
\newline
As a brief summary, our relations with the CHA approach can be described s follows. 

Suppose the  consistency conditions (\ref{081}) are not satisfied for a closed system, not including Observers.
For the CHA  this is the end of the story - for us there is no story, since nothing is perceived. 
We can, however add to the system Observers and their probes, all using the coupling (\ref{A8}). 
Now we can calculate the probabilities, as in Sect.II - but the CHA can calculates the same probabilities, now for a
 larger closed composite {\it  \{ system+probes+memories\}}.

Suppose next  that the conditions  (\ref{081}) are satisfied, the CHA probabilities are available, and there are still no Observers present - this still means little to us.
But if the probes and Observers are  added, the probabilities, calculated in Sect.II for the large composite, will be the same as the ones computed by the CHA, as an Eq.(\ref{081}) for the system only.

In other words, from our point of view, the CHA probabilities coincide with those predicted by the Feynman's rules of Sect II, whenever 
the Observers are present, and are not particularly meaningful in the absence. 

 Finally, since it is up to the Observers to decide which 
 measurements to make, there are many possibilities. For this reason, the CHA cannot single out a particular choice 
 of the projectors $\hat \Pi^\l_{m_\l}$, without any prior knowledge of Observers'  intentions 
 and must, therefore, favour all possible \e{frameworks} in equal measure.
%It is possible that the CHA has other uses, which we failed to mention in our discussion.
The CHA, sometimes deemed to be an {\it interpretation} of quantum theory, has often been criticised 
for the lack of  guidance in choosing a particular \e{physical} representation, in which the calculation of the probabilities
in (\ref{081}) should be made in order to describe the actual experimental occurrences \cite{CHA2}.
With the choices lying with the Observers, no {\it apriori} selection would, of course, be possible. 
%It is  true that the CHA correctly predicts the probabilities, which we already know from Sect...,
%and tests the compatibility of statistical ensembles resulting from bringing in different number of Observers, 
%something we also know from the rules (\ref{01})-(\ref{04}).
%However, its main strength appears to be in making statements about systems, closed to external observations, 
%The utility of such statements is not particularly clear to us at this time. 
%%%%%%%%%%%%%%%%
\section{Conclusions and discussion}
Like every empirical science, quantum mechanics relies on a set of axiomatic rules, which cannot be explained \e{from within the theory}.
The rules need to be consistent, i.e., provide a plausible answer to any question within the theory's area of expertise.
A quantum discourse  often contains issues, which prompt researchers to question its consistency.
%We have tried to deal with them the best we can.  
One such problem is the role and place of a conscious Observer ($O$). Views on the subject vary from assigning the consciousness an active
role \cite{Wig}, to completely excluding the Observer from the narrative \cite{CHA}. Neither of  these extreme positions is particularly appealing.
On one hand, quantum mechanics is a theory by and for intelligent  Observers. On the other hand, as a theory about inanimate physical world \cite{vN},
it is not obliged to provide an insight into the intelligence of its inventors. 
%This suggests that certain half-way might be desirable.
\newline
In search of a compromise, the following analogy may be helpful. Tennis is undoubtably a game by  and for conscious individuals. A ball would  bounce off a wall, 
or off the player's racket.  The latter case is much more complex, since it involves the player, who  sees the ball, takes a decision, 
and makes a deliberate action, all these developments  beyond the reach of classical mechanics. Yet it does not prevent mechanics from calculating the ball's trajectory,
since the only input the theory needs for its predictions, is the force finally exerted on the ball.
%, while the player's deliberations remain 
%safely outside its scope. 
\newline
Thus, one may want to look for something in an act of Observer's perception, which would provide enough input for quantum theory to go on, 
without  making it question the  inner workings of Observer's consciousness. 
%An answer presents itself, yet requires and additional assumption. 
Suppose an act of perception {\it always} results in a change in the state of a material object, destined to carry a record of the perceived outcome (and does not count if no such record is produced). Quantum theory can discuss material objects by assigning to them states 
in an abstract Hilbert space, and specifying their subsequent evolution. Observer's memory may be such a material object, not to be confused
from the Observer's consciousness which, we suspect, quantum theory cannot and should not analyse. Such an assumption does not contradict one's everyday experiences. A person is not continuously aware of facts, and usually needs to consult the memory if asked. If the memory fails, he/she 
may need to consult a note, or a book - another material record, which can be looked at without altering its contents. The analogy is now obvious, although $O$'s memory, which is always at hand,  might enjoy a particularly  intimate relationship with $O$'s consciousness. Placed at the top of the von Neumann chain of devices, which connects $O$'s consciousness  to the outside world, it is at the {\it \e{point  at which we must say 
\e{And this is perceived by the observer}}} \cite{vN}. The details are of no concern to quantum theory, only interested in being able to treat the memory like any other physical object. To complete the analogy, consider a forgetful Observer, who, conscious of his impediment, decides to make a note for himself.
Being a physicist, he takes a spin in a known initial state, and applies a magnetic filed  to prepare it in a state up along  the $z$-axis, if his 
experience was a $yes$, and down the axis, if it was a $no$. The spin is well protected from external influences, and its measurement
along the $z$-axis at a later time, would remind $O$ of what he has seen earlier. The fact that making such a note required certain 
conscious decisions on $O$'s part is of no consequence to quantum theory. If nothing else is changed, the only thing that matters to it is that the spin enters the picture, rotated by different angles in the different version of what happens. 
 %was rotated by a magnetic field. 
\newline
By making  the above assumption, we strike the required balance.
 On one hand, quantum mechanics yields a probability 
of \e{an observer making a certain (subjective) observation}\cite{vN}. On the other hand, a theorist reasoning about what would be seen in an experiment, is able to treat all conscious participants as if they were degrees of freedom describing inanimate objects.
%The fact that the probabilities of Sect... are those of Observer's experiences, and yet the rules of Sect. ... need not take into account the complexities 
%of the individuals' inner life, is reflected in the title of this paper. 
Described in this way, the relationship between Observer's consciousness and his/her memory bears a resemblance to the one between a computer's operating system and its memory stored in in its hard disc,  but only in what relates to his/her experiences of the physical world. An Observer
is free to devise experiments, write poetry, or pray to God - quantum mechanics cannot be a judge of these matters. 
\newline
Evolution of an element of the von Neumann chain, connecting $O$'s memory and the measured system, need not be different
 from any other development, and yet not every interaction constitutes a measurement.
A suitable coupling of a von Neumann's type \cite{vN}, which entangles Observer's probe ($d$) and memory ($\mu$) with the observed system 
($s$) according to  $|\mu_0\ra |d_0\ra |s\ra \to \sum_i c_i|\mu_i\ra |d_i\ra|s_i\ra$, is a particular case of a more general interaction, 
leading to $|\mu_0\ra |d_0\ra |s\ra \to \sum_i \sum_j \sum_k c_{ijk}|\mu_k\ra |d_j\ra|s_i\ra$. In the former special case, the boundary between 
the Observer and the observed system can be moved down the von Neumann chain \cite{vN} including the memory and the probe, and placed 
at the level of the system. It is then possible to say that intermediate observations \e{destroy interference between the system's virtual paths}, in the spirit of Feynman's analysis of the double slit experiment \cite{FeynL}, to which we will return shortly. 
\newline
A valid illustration of the above is the experiment of Sect.V, where an Observer F measures a system using a probe, and later another Observer, W, measures a composite $\{system+F's probe\}$. F may look at his probe, or look away, and what he did would make a difference to what W experiences.
This somewhat surprising result can be explained without granting extra powers to F's consciousness or intelligence, as was suggested, e.g.,  in \cite{Wig}.
If the act of \e{looking} engages F's memory,  this additional degree of freedom must now be included into the calculation of  W's odds.
 At the level of the observed joint system, this amounts to the destruction of interference between the virtual paths in the Hilbert space of the composite, which, in turn, causes
 the disappearance of the interference term, otherwise present in W's probabilities. 
 \newline
There are also broader consequences. Firstly, one would need to assume that the entire body of experimental knowledge 
about physical world is contained in physical records,  e.g.,  in Observer's memories and  notes, and not in his/her consciousness, on the other side
of the \e{observer/observed system divide} \cite{vN}. It follows logically that with all these material records destroyed, all knowledge of what 
actually happened in the physical world would be irretrievably lost. It follows also that in the example of Sect.V,  W's measurement  could in principle engage (difficult though it may be) also F's memory, thus depriving F of the previously gained knowledge about what happen in his experiment.  
However, destruction of F's record by W's measurement does not lead to a formal contradiction. Wigner correctly objected to F's consciousness being in \e{state of suspended animation} \cite{Wig}, 
since is contradicts our experience. This objection is lifted if the discussion is centred on F's memory, which we expect to differ only in complexity 
from any other material object.  
\newline
In his Lectures \cite{FeynL}, Feynman pointed out that a photon, scattered by the electron at one of the slits in a double slit experiment, should  destroy the interference pattern even if it is never detected. Feynman also stressed that many, if not most situations in quantum mechanics are conceptually similar to the double slit example \cite{FeynL}. This is particularly true for our discussion. 
In the scenario $B$  of Sect.V the role of Feynman's photon was played by the Observer F, who, having looked at his probe, carries in his memory 
a record $yes^F/no^F$ of his outcome. This is sufficient for W to find no evidence of  interference in his results, even without F telling him what his outcome was, or even with F having no recollection of the outcome.
Not perfectly isolated from its environment, F's memory may undergo a unitary evolution, so that his $yes^F/no^F$ records evolve into a pair of different orthogonal states, which now include the environment, and as such are no longer recognised by F as valid recollections. 
This does not change W's situation, since his results depend on the presence oft two orthogonal states 
"tagging" the states of the system  [cf. Eq.(\ref{074h})], and not on the nature of these states. 
%If the experiment of Sect... were to take a long time, F could well have no recollection of his 
%outcomes, and W would still see no evidence of  interference in his results. 
\newline
Feynman's example \cite{FeynL} has other interesting consequences.
Suppose that in the experiment of Sect V, half way up F's von Neumann chain (the rest of the chain contains F's retina, neurons, etc.)
a printer prints either $yes^F$ or $no^F$ on a piece of paper. W, an extraordinarily able experimentalist, decides to entangle everything from the 
system to the printout, using a pair of states $[|yes^F\ra \pm |no^F\ra]/\sqrt{2}$, and a probe coupled to whole lab's interior. 
F who is not in the room, which is sealed to the best of W's ability, dedicates himself to evaluating W's odds. 
He reckons that if the isolation of the lab is perfect, W's result will contain an interference term. If, however, a single photon, missed by W, were to strike the printout
on a dark spot and be absorbed, if the printed word is $yes^F$, or on a white spot and be reflected, if the word is $no^F$, 
W's interference term will have to disappear. With several photons left inside, and given a large photon scattering cross section 
of the macroscopic equipment, F could think the second scenario to be more likely, and modify his calculation accordingly.

It is often assumed that quantum theory deals with things so small and delicate, that in any
attempt to probe their condition, the condition is inevitably perturbed. In our previous example, this was not the case. A small and delicate 
photon appears to affect the state of something large, classical and fairly robust. This brings us to the second topic of our discussion.  
There is also a controversy surrounding the role and status of the quantum wave function, which stems from the desire to see the outcomes of an experiment in progress as 
a reflection on the real-time evolution of certain physical state (or substance), associated with the observed system. 
This view was broadly outlined in Sect.III.
%Often made  parallel between the wave function and a \e{physical} electromagnetic field is not particularly helpful. In general, a solution of the Maxwell's equations is itself a
%kind of  photon's  wave function, which becomes a real  \e{classical} field only for large occupation numbers \cite{FeynS}.
Moreover, anyone wishing to make unitary evolution the only basic principle of quantum theory
%Eqs.(\ref{011})-(\ref{031})
 %of Sect... to be the basic principles of quantum theory,
  immediately meets with 
the problem of \e{collapse} of a quantum state, be it of a pure, or of a mixed kind. 
A sudden decimation of the wave function after an Observer obtains a definite outcome cannot be described by the Schroedinger equation, 
and requires an additional \e{projection postulate} \cite{vN}. Accepting that a collapse is a physical phenomenon, prompts further 
questions about when and how exactly it occurs.
One wishing to avoid these questions by preserving the integrity of the wave function against all odds, 
may decide to send its unused bits to parallel universes, as happens, for example, in the many-worlds interpretation of quantum mechanics \cite{MW}. A comparative analysis of these two viewpoints can be found, for example, in \cite{Deu}.
\newline
We argue that the above problems are artificial. They need not be solved, but can rather be dismissed once the appropriate terminology is adopted. 
It is possible to admit the rules of Sect.II. as a basic principle, and consider Eqs.(\ref{011})-(\ref{031}) of Sect.III to be their derivable consequences,
serving mostly to simplify the calculations. 
Now Eqs.(\ref{01})-(\ref{04}) are but a tool used by Alice, not taking part in the experiment herself, to reason about the odds on the outcomes
perceived by the $L$ intelligent Observers, were this experiment to be performed.
Alice knows that the $\l$-th
Observer's inquiry about the system is represented by an operator $\hat Q^\l$, 
and associates with the system a Hamiltonian $\h^S$.
At the end of each run a sequence of Observer's outcomes would be recorded, the numbers of identical records counted, 
and used to measure the probabilities which Alice is calculating in the comfort of her office. While doing so, Alice is not worried about the Observers' 
consciousnesses, since in her calculation each participant is represented by the degrees of freedom of his/her material memory. 
It does not matter to Alice whether the Observers communicate with each other (provided their memories do not form a part 
of the measured system). 
%or exchange their results during the course of the experiment.
Her job is to evaluate matrix elements of a unitary operator, $\u(t_\l-t_{\l-1})=\exp[-i\h(t_\l-t_{\l-1})]$, between the eigenstates of the 
$\hat Q^\l$, representing the Observers' measurements, multiply them, add the products as appropriate, and square the moduli of the resulting  complex numbers. There is no mention of  a wave function, expected to evolve in a continuous manner, nor any need to look for a home for its unwanted
parts. However, Alice may have noticed that sometimes the formulae of Sect.III offer an easier way of calculating the probabilities, 
than the just described basic procedure. For example, instead of calculating the odds for $L$ Observers she can evaluate, 
as we did in Sect.X, the probabilities for a single Observer plus $L-1$ probes, already coupled, but as yet unregistered. 
This involves matrix elements of a single evolution operator (\ref{073h})  and is, for this reason, a simpler task. In the process, Alice may begin to put more faith in
the universal value of unitary evolution, but does not have to do so. The rules of Sect. II serve only to establish statistical correlation between 
at least two Observers' experiences, and cannot be reduced further \cite{DSepl1}. In the above example, the $L$-th Observer's definite outcome 
would \e{ collapse} the state, making Alice wonder about the destiny of the rest of the so far unitary evolved wave function. 
But, as we said, this is by no means necessary. Alice could as well proclaim \e{ the experiment finished, the desired probability evaluated}
[cf. \cite{FeynL}], and close her notebook. She would refuse to answer Bob's question \e{what happened to the system after that?}
But if asked instead \e{what would be the results of an $L+1$-st  Observer, who decides to join the experiment at some $t_{L+1}>t_L$?},
she would reopen the notebook, make a new calculation for the {\it entire} new series of outcomes, $\{Q^{L+1}_{m_{L+1}}...\gets Q^{\l}_{m_{\l}}....\gets Q^0_{m_0}\}$, and then close it again. 
\newline
In summary, we found elementary quantum mechanics consistent, in the sense of being able to provide an unambiguous answer at least in  the hypothetical situations considered in this work. The  \e{minimalist} view \cite{DSepl1}, advocated here, comes at a price of making certain additional assumptions. 
In particular, the theory is deemed to predict statistical correlations between at least two of the Observer's {\it \e{subjective observations}} \cite{vN}, 
accompanied by producing, or consisting in consulting's a record in the Observer's material memory. 
With the line between Observer's consciousness and the physical world drawn at the memory's level, Feynman's {\it general principles} \cite{FeynL}
need to be applied to the {\it entire} duration of the experiment. The focus is thereby shifted from a continuously 
evolving wave function to the transition amplitudes (\ref{01})-(\ref{02}), seen as mere tools of human reasoning. This helps one avoid unfruitful (in our opinion) discussions about 
the exact moment in which a quantum state \e{collapses} \cite{Deu}, or whether the unused parts of the state found their use in parallel worlds \cite{MW}.
\newline
Expected restrictions on potential applications of quantum theory are also considerable. Quantum mechanics is not expected to make statements  about human consciousness, and cannot explain how consciousness addresses the memory, or retrieves the memorised information from it. 
With the probabilities referring to humans experiences (actual, or possible {\it in principle}), further questions arise about the theory's 
retroductive powers (see e.g., \cite{CHA3}), as well as about such global concepts as the wave function of the entire Universe. 
\newline
We will return to these issues in our future work, and conclude with a picture, in which various Alices and Bobs perform experiments of their choice, perceive the outcomes, 
% leave notes, 
 memorise and forget, produce and destroy records of their outcomes, wittingly or unwittingly, and share or not their experiences with each other. 
In the meantime, Carols (the roles can of course be exchanged)  are evaluating the likelihoods of Alices' and Bobs' outcomes, taking into account 
only the changes their actions may produce in the inanimate physical world.
%on the assumptions about the experimental setup. An everybody is having fun!
%%%%%%%%%%%%%%%%%%
\section{Appendix. Coupling a probe to the system.}
Consider a system $S$ in a Hilbert space of a dimension N, and an operator with $M\le N$ distinct eigenvalues $\tilde S_m$, 
($\Delta(x-y)=1$, if $x=y$, and $0$ otherwise)
\begin{eqnarray} \label{A1}
\hat S =\sum_{n=1}^N |s_n\ra S_n\la s_n|=\sum_{m=1}^{M_L}\S_m  \sum_{n=1}^N \Delta(\S_m -\la s_n|\hat S| s_n\ra) 
|s_n\ra \la s_n|\equiv \sum_{m=1}^{M_L}\S_m\hat \pi_m.
\end{eqnarray}
Consider also a probe (a von Neumann's pointer), a  massive particle in one dimension, with a coordinate $q$, and a momentum operator ($\hbar=1$)
$\hat p$ $\la q'|\hat p|q\ra= -i\delta(q-q') \partial_q$. The Hamiltonian, coupling the system to the pointer will be $\hat H_{int} =\hat p \hat S\delta(t-\tau)$, so that the evolution operator over a period $\tau-\epsilon <t < \tau+\epsilon$ is $\u(\tau+\epsilon,\tau-\epsilon) 
= \exp(-i\hat p \hat S)$ 
%whose matrix elements are given by 
%\begin{eqnarray} \label{A2}
%\la q'|\la s_{n'}|\u(\tau+\epsilon,\tau-\epsilon)|q\ra |s_n\ra = \delta_{nn'} \delta(q'-q-S_n), 
%\end{eqnarray}
is given by
\begin{eqnarray} \label{A3}
\u(\tau+\epsilon,\tau-\epsilon)=
%\sum_{n,n'=1}^N\int dq dq' |q'\ra|s_{n'}\ra\la q'|\la s_{n'}|\u(\tau+\epsilon,\tau-\epsilon)|q\ra |s_n\ra
%\la q|\la s_n|=\n
\sum_{n=1}^N \int dq  |q+S_n\ra |s_n\ra\la s_n|\la q|.
\end{eqnarray}
Initially, the system and the pointer are described by a  product state $|\Phi_0\ra =|s\ra|G\ra$, where
\begin{eqnarray} \label{A4}
|s\ra =\sum_{n=1}^N c_n|s_n\ra, \q \text{and}\q |G\ra= \int dq G(q)|q\ra, 
\end{eqnarray}
so that
\begin{eqnarray} \label{A4}
\u(\tau+\epsilon,\tau-\epsilon)|\Phi_0\ra = \sum_{n=1}^N c_n|s_n\ra |G(S_n)\ra =\q\q\q\q\q\q\q\q\n
\sum_{m=1}^M |G(\S_m)\ra \sum_{n=1}^N \Delta(\S_m -\la s_n|\hat S| s_n\ra)  c_n|s_n\ra
= \sum_{m=1}^M |G(\S_m)\ra \pi_m |s\ra.
\end{eqnarray}
where $|G(Z)\ra \equiv \int dq G(q-Z)|q\ra$. 
Let $G(q)$ be a Gaussian of a width $\Delta q$, 
\begin{eqnarray} \label{A5}
G(q)= C \exp(-q^2/\Delta q^2), \q \int |G(q)|^2 dq=1,  
\end{eqnarray}
and send $\Delta q$ to zero, so that $\la G(\S_{m'}) |G(\S_m)\ra \to \delta_{mm'}$. Although the probe's Hilbert space 
has of infinite dimensions, we will only need its $M+1$ orthogonal states,  
\begin{eqnarray} \label{A6}
|d_0\ra \equiv |G\ra,\q \text{and} \q  |d_j\ra \equiv |G(\S_j)\ra, \q j=1,...M. 
\end{eqnarray}
and will describe application of the coupling (\ref{A3}) by saying that 
{\it \e{ the system is coupled to (entangled with) the probe according to}}
\begin{eqnarray} \label{A7}
|d_0\ra |s\ra \to \sum_{m=1} ^M |d_m\ra \hat \pi_m|s\ra. 
\end{eqnarray}
\newline
The coupling (\ref{A3}) can be reversed by applying $-\hat H_{int}$,  whose action is defined as 
\begin{eqnarray} \label{A8}
 |d_j\ra \hat \pi_j|s\ra\to   |d_0\ra \hat \pi_j|s\ra.
\end{eqnarray}
\begin{center}
\textbf{Acknowledgements}
\end{center}
Financial support of
MCIU, through the grant
PGC2018-101355-B-100(MCIU/AEI/FEDER,UE)  and the Basque Government Grant No IT986-16.
is acknowledged by DS.

\end{document}